\RequirePackage{fix-cm}	

\documentclass[smallextended, natbib, final]{svjour3}      

\smartqed  
\usepackage{graphicx}
\usepackage[misc]{ifsym}
\usepackage[english]{babel}
\usepackage{amsmath} 
\usepackage{amssymb}
\usepackage{booktabs} 
\usepackage{dsfont}
\usepackage[]{hyperref}
\usepackage{pgfplots}
\usepackage[per-mode=fraction]{siunitx}
\usepackage{tikz}
\usepackage{csquotes}
\usepackage[all]{nowidow} 
\usepackage[font=small]{caption}
\usepackage[font=small]{subcaption} 
\usepackage{bm}
\usepackage{setspace}
\usepackage{mathtools}
\usepackage{multirow}
\usepackage{pifont}
\usepackage{xcolor}

\newcommand*{\vm}[1]{\boldsymbol{#1}} 
\newcommand*{\trans}{^{\mathsf{T}}}

\newcommand*{\ex}[1]{\mathds{E}\left[#1\right]}

\newcommand*{\Ex}{\mathds{E}} 
\newcommand*{\cov}{\text{Cov}}
\newcommand*{\R}{\mathbb{R}}
\newcommand*{\dd}{\mathrm d}

\newcommand*{\func}{\psi}
\newcommand*{\inp}{\xi}

\newcolumntype{M}[1]{>{\centering\arraybackslash}m{#1}}

\usetikzlibrary{arrows.meta}
\usetikzlibrary{decorations.markings}
\usetikzlibrary{intersections, pgfplots.fillbetween}
\usetikzlibrary{shapes}
\usetikzlibrary{shapes.misc}
\usetikzlibrary{positioning}
\usetikzlibrary{backgrounds}
\pgfplotsset{compat=1.18}
\usepgfplotslibrary{groupplots}
\definecolor{blue1}{rgb}{0.117647,0.250980,0.705882}
\definecolor{red1}{rgb}{0.898039,0.125490,0.125490}
\definecolor{green1}{rgb}{0.250980,0.745098,0.250980}
\definecolor{gray1}{rgb}{0.85,0.85,0.85}

\journalname{}
\begin{document}

\title{A software framework for stochastic model predictive control of nonlinear continuous-time systems (GRAMPC-S)}

\titlerunning{A software framework for stochastic model predictive control ...}

\author{Daniel Landgraf \and Andreas V{ö}lz \and Knut Graichen}

\institute{Daniel Landgraf (\Letter)\at \email{daniel.dl.landgraf@fau.de}
\and
Andreas V{ö}lz \at \email{andreas.voelz@fau.de}
\and
Knut Graichen \at \email{knut.graichen@fau.de} \\\\
Chair of Automatic Control, Friedrich-Alexander-Universität Erlangen-Nürnberg,
Cauerstraße 7, 91058 Erlangen, Germany\newline
}
\date{Received: date / Accepted: date}

\maketitle
\begin{abstract}
This paper presents the open-source stochastic model predictive control framework GRAMPC-S for nonlinear uncertain systems with chance constraints. It provides several uncertainty propagation methods to predict stochastic moments of the system state and can consider unknown parts of the system dynamics using Gaussian process regression. These methods are used to reformulate a stochastic MPC formulation as a deterministic one that is solved with GRAMPC. The performance of the presented framework is evaluated using examples from a wide range of technical areas. The experimental evaluation shows that GRAMPC-S can be used in practice for the control of nonlinear uncertain systems with sampling times in the millisecond range.
\keywords{Stochastic model predictive control \and Nonlinear model predictive control \and Stochastic nonlinear dynamics \and Chance constraints \and Real-time implementation}
\end{abstract}

\section{Introduction} \label{sec:Introduction}

Model predictive control (MPC) is an advanced control method that has been used with great success for the control of complex systems~\citep{MAYNE20142967, bib:camacho2003}. It is based on the solution of an optimal control problem~(OCP) on a moving horizon and can be applied to nonlinear systems with state and input constraints. A model of the system is used to predict the states up to a prediction horizon and to optimise the control input accordingly. The numerical effort that is required to solve the OCP, however, is a major challenge for the real-time application of MPC. Therefore, suboptimal approaches have been developed that significantly reduce the computation time~\citep{Diehl.2005,Graichen.2010, Richter:2012}.

MPC requires a sufficiently accurate model of the system dynamics, as otherwise the control performance decreases and the satisfaction of constraints is no longer guaranteed. Most systems cannot be modeled exactly, with the result that system models are often based on simplifications which lead to uncertainty about the system behaviour. This systematic uncertainty is referred to as epistemic uncertainty and can appear, for example, in the form of uncertain system parameters. It is to be distinguished from aleatoric uncertainty, which is inherent in the system behaviour. The system dynamics cannot be modeled by ordinary differential equations (ODE) in such a case, but must be represented by stochastic differential equations (SDE)~\citep{Nisio2015}. The deterministic formulation of the OCP in MPC, however, does not allow to consider these uncertainties systematically, which necessitates extensions of the MPC concept.

One way to consider uncertainties in control is robust model predictive control (RMPC), which assumes that the uncertainties lie in a bounded set~\citep{Bemporad.1999}. \citet{Campo.1987} and~\citet{Allwright.1992} developed the robust min-max MPC, where the constraints must be satisfied for all realizations of the uncertainty and the cost function is evaluated for the worst case. However, this approach can lead to excessively conservative results or even to infeasible problems. Tube-based robust MPC mitigates these disadvantages by defining a parameterized control law whose parameterization is used as optimization variable rather than the control input~\citep{Langson.2004,Mayne.2005,Rakovic.2012}. \citet{Zeilinger.2014} show that RMPC can be executed in real time with guarantees on feasibility and stability for linear systems. 

The disadvantages of RMPC are its design for the worst case scenario and that no prior knowledge of the stochastic properties of the uncertainties can be taken into account. Stochastic model predictive control (SMPC), in contrast, considers state variables of the system as random variables with associated probability density functions \citep{Mesbah.2016,Kouvaritakis.2016}. Constraints are formulated in the form of chance constraints, which must be satisfied with a certain probability. The evaluation of chance constraints requires the knowledge of the probability density function of the predicted states, which necessitates the propagation of uncertainties through the system dynamics. While this is simple for linear systems with Gaussian uncertainty distributions, it is a major challenge for nonlinear systems, which necessitates the use of approximative propagation methods~\citep{Dunik2020,Landgraf.2023}. In particular, successive linearization of the system dynamics is often used for systems with weak nonlinearities, enabling the use of SMPC methods for linear systems~\citep{Cannon.2009,Ma.2015}. However, the linearization of the system functions results in errors that are avoided if the nonlinear system dynamics are used for the uncertainty propagation. Sampling-based methods represent the uncertainties by a set of sampling points and evaluate the nonlinear system dynamics at these points. The samples can be selected either with deterministic transformations, such as the unscented transformation~\citep{Julier2004Unscented, Volz.2015}, or by Monte Carlo sampling~\citep{Maciejowski.2007, Kantas.2009}. Furthermore, several SMPC approaches exist based on polynomial chaos expansion, see e.g. \citet{Paulson.2014}, \citet{Paulson.2018}, and \citet{Fagiano.2012}.

One of the main problems of MPC and SMPC is the computational effort involved in solving the OCP. The application in real-time therefore requires the implementation of efficient algorithms in software frameworks. Various toolboxes are available for deterministic systems, such as acados \citep{Verschueren.2021}, GRAMPC \citep{Englert.2019}, GRAMPC-D \citep{burk2022}, FalcOpt~\citep{FalcOpt_Theory} or VIATOC~\citep{Kalmari.2015}. However, these cannot be used for stochastic systems, as they are not directly able to predict state uncertainties. To overcome this limitation, specific SMPC toolboxes have been published. \citet{Gonzalez.2020} developed a Matlab toolbox for linear systems with additive noise that can consider both chance constraints and random scenarios. \citet{Brudigam.2023} also published a Matlab toolbox for SMPC of linear systems with chance constraints, which contains a number of examples as well as the possibility to switch between stochastic and deterministic approaches. Moreover, \citet{Andrian.2019} presented a Matlab SMPC toolbox, which calculates the uncertainties using polynomial chaos expansion (PCE), which is suitable for linear systems. Furthermore, PCE toolboxes for Matlab \citep{Petzke.2020} and Julia \citep{Muhlpfordt.2020} have been published that can be used for nonlinear systems. Further nonlinear SMPC toolboxes are UKF-SNMPC~\citep{Bradford.2018}, where both the state estimation and the propagation of uncertainties are executed using the uncented transformation, and Bioptim~\citep{Michaud.2023}, which is a recently published SMPC toolbox for biomechanics written in Python. However, the focus of these toolboxes is not on the computational efficiency that would be required to control highly dynamic systems in real time. Moreover, most frameworks consider only one uncertainty propagation method or are limited to specific system classes, such as linear systems.

This paper presents the open-source software framework GRAMPC-S for SMPC of nonlinear systems, which can be used for real-time control with sampling times in the millisecond range. It provides C++ implementations of several propagation methods for stochastic uncertainties as well as deterministic approximations of chance constraints. The stochastic uncertainty can be propagated either by linearization of the system dynamics, by several sampling based methods or by PCE. The choice of the propagation method represents a trade-off between approximation accuracy and computational effort. While some methods such as Taylor linearization and unscented transformation lead to low computation time, other approaches such as PCE and Monte-Carlo sampling can approximate the propagation problem more accurately. The approximative uncertainty propagation is applied to transform the stochastic OCP into a deterministic one, which is solved with the MPC toolbox GRAMPC~\citep{Englert.2019}. Uncertainties can be considered in the form of uncertain initial states, e.g. due to measurement noise or state estimation, uncertain system parameters or an additive Wiener process in the system dynamics. In addition, GRAMPC-S can consider Gaussian processes that are learned from measurement data to represent unknown parts of the system dynamics. The variance of the Gaussian process is evaluated and explicitly taken into account by the controller. The structure of the framework allows to extend the code by further propagation methods, probability density functions or kernels for Gaussian process regression. Besides the usage as C++ code, GRAMPC-S can be integrated into Matlab/Simulink, which also allows for code porting, e.g. to dSPACE hardware. Since the focus of this paper is on the presentation of GRAMPC-S, the uncertainty propagation methods are only briefly outlined. A more detailed overview of the propagation methods can be found in \citet{Landgraf.2023}.

The paper is structured as follows. Section~\ref{sec:problem_formulation} outlines the stochastic OCP and the associated problem formulation. Section~\ref{sec:stoch_MPC} describes the SMPC framework including the uncertainty propagation, the consideration of chance constraints, the deterministic reformulation of the stochastic OCP, and the consideration of Gaussian processes. Section~\ref{sec:structure} presents the structure of the software framework and its utilization. Section~\ref{sec:evaluation} evaluates the performance of GRAMPC-S with several simulation examples and validates its practical applicability for a laboratory-scale experiment. Section~\ref{sec:conclusion} concludes the paper.

The following notations are used in this paper. A vector $\vm x$ is written in bold. The expected value of a random variable $\vm x$ is denoted by $\mathds{E}\left[ \vm x \right]$ and the covariance matrix by $\text{Cov} \left[ \vm x \right] = \mathds{E}\big[ \left(\vm x - \mathds{E}\left[ \vm x\right] \right)  \left(\vm x - \mathds{E}\left[ \vm x\right] \right)\!\vphantom{x}\trans \big]$. The cross-covariance matrix of two random variables $\vm x$ and $\vm y$ is denoted by $\text{Cov}\left[ \vm x ,\, \vm y \right] = \mathds{E}\big[ \left(\vm x - \mathds{E}\left[ \vm x\right] \right)  \left(\vm y - \mathds{E}\left[ \vm y\right] \right)\!\vphantom{x}\trans \big]$ and the variance of a univariate random variable $x$ is denoted by $\text{Var}[x]$. A Gaussian distributed random variable $\vm x$ with mean $\vm \mu $ and covariance matrix $\vm \Sigma$ is denoted by $\vm x \sim \mathcal{N}\left(\vm \mu,\, \vm \Sigma \right)$ and a univariate uniformly distributed random variable $y$ with lower bound $a$ and upper bound $b$ is denoted by $y \sim \mathcal{U}\left(a,\,b\right)$. The probability of an event $A$ is referred to as $\mathds{P}\left[A\right]$. The inner product $\left\langle f, g \right\rangle$ of two functions $f: \mathbb{R} \rightarrow \mathbb{R}$ and $g: \mathbb{R} \rightarrow \mathbb{R}$ is defined by the integral $\int_{x_1}^{x_2} f(x) g(x) w(x) \,\text{d}x$ with a weight function $w : \left[x_1, \, x_2 \right] \rightarrow \mathbb{R}$ that must be specified. For reasons of readability, the explicit dependency on time $t$ may be omitted for time-dependent variables and the time derivative is written as $\dot{\vm x}(t) = \frac{\text{d} \vm x(t)}{\text{d} t}$.
 \clearpage
\section{Problem formulation} \label{sec:problem_formulation}
SMPC is based on the iterative solution of a stochastic OCP, where not only the states but also the system parameters and parts of the system dynamics can be uncertain. The most general OCP formulation that can be considered with GRAMPC-S is
\begin{subequations} \label{eq_ocp}
\begin{align}
 \min_{\vm u} \,\,\quad& J\left( \vm u; \, \vm p, \, \vm x_0 \right) = \mathds{E} \left[ V(\vm x(T), \, \vm p) + \int\limits_0^T l(\vm{x}, \vm{u}, \, \vm p) \, \text{d}t \right] \label{eq_ocp_cost}\\
 \operatorname{s.t.} \quad\;\,& \text{d}\vm{x} = \vm f(\vm{x}, \, \vm{u}, \, \vm p) \, \text{d}t + \vm \sigma_w \, \text{d} \vm w \,,\quad \vm{x}(0) = \vm x_0 \label{eq_ocp_dyn}\\
 & \mathds{P} \left[h_i(\vm{x}, \, \vm{u}) \leq  0 \right] \geq \alpha_i \,,\qquad \hspace{0.75cm} i = 1, \dots, N_h \label{eq_ocp_con}\\
 & \mathds{P} \left[h_{T,\,j}(\vm{x}(T)) \leq  0 \right] \geq \alpha_{T,\,j} \,, \quad \hspace{0.23cm}\, \, j = 1, \dots,  N_{h_T} \label{eq_ocp_conTerm}\\
 & \vm u_\text{min} \leq \vm{u} \leq \vm u_\text{max}  \label{eq_ocp_inputCon}
\end{align}%
\end{subequations}%
with state $\vm x \in \R^{N_x}$, control $\vm u \in \R^{N_u}$, system parameters $\vm p \in \R^{N_p}$, and prediction horizon $T \in \R$. The initial state $\vm x_0$ and the parameters $\vm p$ are assumed to be random variables with known probability density functions. 
The cost functional $J$ in \eqref{eq_ocp_cost} consists of the expected value of the  integral cost $l$ and the expected value of the terminal cost $V$. The system dynamics is given by the SDE \eqref{eq_ocp_dyn}, where $\vm f$ describes the system dynamics function and $\vm w \in \R^{N_x}$ is a Wiener process with diffusion term $\vm \sigma_w \in \R^{N_x \times N_x}$. The probability distribution of $\vm x$ is predicted by integrating the system dynamics, with the initial state $\vm x_0$ representing the estimated or measured state in the current time step. In addition, the probabilistic inequality constraints \eqref{eq_ocp_con} with constraint functions $ h_i$ and $i = 1, \dots, N_h$, must be satisfied with probabilities $0 \leq \alpha_i \leq 1$ and the terminal inequality constraints \eqref{eq_ocp_conTerm} with terminal constraint functions $h_{T,j}$ and $j = 1, \dots, N_{h_T}$ must be satisfied with probabilities $0 \leq \alpha_{T,\,j} \leq 1$. Moreover, the box constraint \eqref{eq_ocp_inputCon} can be taken into account for the control input $\vm u$.

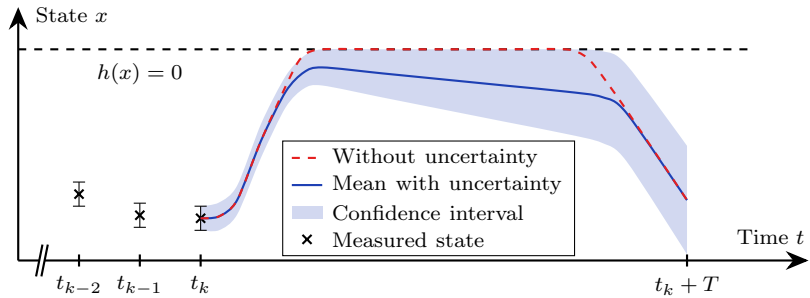
\begin{figure}
	\centering
	\def\MarkLt{6pt}
\def\MarkSep{2pt}

\tikzset{cross/.style={cross out, draw=black, minimum size=2*(#1-\pgflinewidth), inner sep=0pt, outer sep=0pt}, cross/.default={2.5pt}}

\tikzset{
  marks/.style={
    postaction={decorate,
      decoration={
        markings,
        mark=at position #1 with
          {
              \begin{scope}[xslant=0.2]
              \draw[line width=\MarkSep,white,-] (0pt,-\MarkLt) -- (0pt,\MarkLt) ;
              \draw[-] (-0.5*\MarkSep,-\MarkLt) -- (-0.5*\MarkSep,\MarkLt) ;
              \draw[-] (0.5*\MarkSep,-\MarkLt) -- (0.5*\MarkSep,\MarkLt) ;
              \end{scope}
          }
       }
    }
  },
  TwoMarks/.default={0.5},
}

\begin{tikzpicture}[scale = 0.8]
\draw[-{Stealth[scale=1.3,angle'=45]},thick, marks=0.03] (-2,1) -- (11, 1) node[above= 0.9mm, xshift=-5mm] {\small Time $t$};
\draw[-{Stealth[scale=1.3,angle'=45]},thick] (-2,1) -- (-2, 5.2) node[right= 1mm, yshift=-1mm,minimum height=6mm] {\small State $x$};
\draw[dashed, thick] (-2,4.5) -- (10,4.5);

\draw[thick] (-1, 1.1) -- (-1, 0.9) node[below] {\small $t_{k-2}$};
\draw[thick] (0, 1.1) -- (0, 0.9) node[below] {\small $t_{k-1}$};
\draw[thick] (1, 1.1) -- (1, 0.9) node[below] {\small $t_k$};
\draw[thick] (9, 1.1) -- (9, 0.9) node[below] {\small $t_k + T$};

\draw [name path=A, draw=none]  plot[smooth] coordinates {(1,1.9) (1.3,1.96) (1.6,2.24) (2,3.28) (2.4,4.06) (2.7,4.39) (3,4.5) (4,4.5) (5, 4.5) (6, 4.5) (7, 4.5) (7.6, 4.47) (8, 4.25) (9,2.9)};

\draw [name path=B, draw=none]  plot[smooth] coordinates {(1,1.5) (1.3,1.5) (1.6,1.76) (2,2.72) (2.4,3.54) (2.7,3.85) (3,3.9) (4,3.7) (5, 3.5) (6, 3.3) (7, 3.1) (7.6, 2.95) (8, 2.65) (9,1.1)};

\tikzfillbetween[of=A and B]{blue1, opacity=0.2};

\draw (-1,2.1) node[cross,black, thick] {};
\draw (0,1.75) node[cross,black, thick] {};
\draw (1,1.7) node[cross,black, thick] {};

\draw[black] (1, 1.5) -- (1, 1.9);
\draw[black] (0, 1.55) -- (0, 1.95);
\draw[black] (-1, 1.9) -- (-1, 2.3);

\draw[black] (0.9, 1.5) -- (1.1, 1.5);
\draw[black] (0.9, 1.9) -- (1.1, 1.9);
\draw[black] (-0.1, 1.55) -- (0.1, 1.55);
\draw[black] (-0.1, 1.95) -- (0.1, 1.95);
\draw[black] (-1.1, 1.9) -- (-0.9, 1.9);
\draw[black] (-1.1, 2.3) -- (-0.9, 2.3);

\draw[thick, blue1] plot[smooth] coordinates {(1,1.7) (1.3,1.73) (1.6,2) (2,3) (2.4,3.8) (2.7,4.12) (3,4.2) (4,4.1) (5, 4.0) (6, 3.9) (7, 3.8) (7.6, 3.71) (8, 3.45) (9,2)};

\draw[thick, red1, dashed] plot[smooth] coordinates {(1,1.7) (1.3,1.73) (1.6,2) (2,3) (2.5,4.1) (2.7,4.4) (3,4.5) (4,4.5) (5, 4.5) (6, 4.5) (7, 4.5) (7.3, 4.4) (7.5, 4.2) (8, 3.5) (9,2)};

\draw[draw=black] (2.35,1.08) rectangle ++(4.8,1.9);

\node[text width=35mm] at (5.35,2.7) {\small Without uncertainty};
\node[text width=35mm] at (5.35,2.25) {\small Mean with uncertainty};
\node[text width=35mm] at (5.35,1.8) {\small Confidence interval};
\node[text width=35mm] at (5.35,1.35) {\small Measured state};

\draw [thick, red1, dashed] (2.5, 2.7) -- (3, 2.7);
\draw [thick, blue1] (2.5, 2.25) -- (3, 2.25);
\fill [blue1, opacity=0.2] (2.5,1.7) rectangle ++(0.5,0.2);
\draw (2.75,1.35) node[cross,black, thick] {};

\node at (0,4.1) {\small $h(x) = 0$};

\end{tikzpicture}
	\caption{Illustrative example for the solution of a constrained OCP with and without considering uncertain system dynamics.}
	\label{fig:SMPC_concept}
\end{figure}

The stochastic OCP~\eqref{eq_ocp} differs from the classical deterministic MPC setting by the presence of uncertainties. Even if the system dynamics are known exactly, an uncertain initial state $\vm x_0$ or uncertain system parameters $\vm p$ will result in uncertain predicted states. Figure~\ref{fig:SMPC_concept} illustrates the effect of uncertainties on the solution of an OCP with one state constraint. In the deterministic case, the predicted state is allowed to reach the boundary of the inequality constraint $h(x) = 0$. If the system dynamics are uncertain, the probabilistic constraint requires a deviation between the mean of the state and $h(x) = 0$. The uncertainty in the system dynamics often results in an increase of the state uncertainty over the prediction horizon. Therefore, the distance between the mean of the predicted state and the boundary $h(\vm x) = 0$ must increase accordingly, in order to satisfy the probabilistic constraint. Note that the uncertainty of the state does not necessarily need to increase with prediction time, e.g. it decreases for stable linear systems.

\section{Stochastic model predictive control} \label{sec:stoch_MPC}
A stochastic OCP cannot be solved directly using MPC solvers, for which reason the SMPC framework presented in this paper transforms the stochastic OCP into a deterministic OCP. The proposed approach requires three parts: The propagation of uncertainties through a nonlinear function as it appears in \eqref{eq_ocp_dyn}, the approximation of the chance constraints \eqref{eq_ocp_con} and \eqref{eq_ocp_conTerm} and the reformulation of the OCP. Section~\ref{sec:uncert_prop} shortly describes the methods of uncertainty propagation implemented in GRAMPC-S. Section~\ref{sec:chance_constr} presents different ways of reformulating chance constraints as deterministic constraints. Section~\ref{sec:det_reformulation} shows how the uncertain states and system parameters as well as the Wiener process $\vm w$ can be represented and taken into account by the stochastic model predictive controller. Section~\ref{sec:GP} outlines the consideration of Gaussian processes in the system dynamics.

\subsection{Uncertainty propagation} \label{sec:uncert_prop}
SMPC requires to predict the states of the system \eqref{eq_ocp_dyn} up to the prediction horizon. Since the states are random variables, this necessitates not only predicting the mean of the states, but also calculating their stochastic distribution. Formally, the temporal evolution of the probability density functions of the states can be described by the Fokker-Planck partial differential equation \citep{Challa2000,gardiner2009}. However, the Fokker-Planck equation cannot be solved analytically for general nonlinear system dynamics, but only for some special cases \citep{Luo2017}.

A further approach is to predict only a finite number of stochastic moments, such as mean and variance. In this case, the problem of uncertainty propagation can be defined as follows. Given a probability density function $p(\vm \inp)$ of a random variable $\vm \inp \in \R^{N_\inp}$, calculate a finite number of stochastic moments of an output variable
\begin{align} \label{eq:uncert_prop_problem}
	\vm y = \vm \func(\vm \inp) \in \R^{N_y}
\end{align}
with the nonlinear function $\vm \func: \R^{N_\inp} \rightarrow \R^{N_y}$.
In this section, the uncertainty prediction for general nonlinear functions $\vm  \func $ with inputs $\vm \inp$ is presented. For readability reasons, the mean and the covariance matrix of $\vm \inp$ are denoted as $\vm \mu_\inp  =\ex{\vm \inp}$ and $\vm \Sigma_\inp = \text{Cov}\left[\vm \inp\right]$. The propagation methods approximate the mean and the covariance matrix of $\vm y$, where the estimated values are denoted accordingly as $\hat{\vm \mu}_y \approx \ex{\vm y}$ and $\hat{\vm \Sigma}_y \approx \text{Cov}\left[ \vm y \right]$. A detailed overview and a comparison of the propagation methods is presented in \citep{Landgraf.2023}.
The application of the uncertainty prediction in SMPC is discussed subsequently in Section \ref{sec:det_reformulation}.

\subsubsection{Taylor linearization}
The propagation of uncertainty is considerably complicated by the fact that the propagation function $\vm \func$ in~\eqref{eq:uncert_prop_problem} can be nonlinear. A simple way to avoid this problem is to approximate it around a point $\vm \mu_\inp$ using the first-order Taylor series approximation
\begin{align}
	\vm{\tilde \func}(\vm \inp) = \vm \func(\vm \mu_\inp) + \frac{\partial \vm \func}{\partial \vm \xi}(\vm \mu_\inp) (\vm \inp - \vm \mu_\inp)\,, \label{eq:taylor_linear}
\end{align}
where $\frac{\partial \vm \func}{\partial \vm \xi}(\vm \mu_\inp)$ denotes the Jacobian of $\vm \func$ at the mean of $\vm \inp$. Using this linearization, the mean and covariance matrix of $\vm y$ can be calculated with \citep{Chui2009Kalman,Lewis1986Optimal,Simon2006Optimal}
\begin{subequations} \label{eq_lin_Taylor_mean_and_cov}
\begin{align}
	\vm{\hat \mu}_y &= \vm \func(\vm \mu_\inp) \label{eq:mean_taylor_linear}\\
	\vm{\hat \Sigma}_y &= \frac{\partial \vm \func}{\partial \vm \xi}(\vm \mu_\inp) \vm \Sigma_\inp \frac{\partial \vm \func}{\partial \vm \xi}(\vm \mu_\inp)\trans \label{eq:covariance_taylor_linear} \,,
\end{align}
\end{subequations}
which is used in the same way in the prediction step of the extended Kalman filter. The first-order Taylor series approximation is a simple uncertainty propagation method that assumes that the function $\vm \func$ is differentiable, the random variable $\vm \inp$ is described sufficiently accurately by its mean and covariance matrix, and the linearization is a sufficiently accurate approximation of the nonlinear function $\vm \func$. Propagation functions that are strongly nonlinear within the support of $\vm \inp$ result in increasing errors of the linearization~\eqref{eq:taylor_linear} and therefore of the estimated mean and covariance matrix in~\eqref{eq_lin_Taylor_mean_and_cov}. Beside the linearization, also higher orders of the Taylor series approximation can be used for uncertainty propagation, e.g. explicit formulas for the calculation of $\vm{\hat \mu}_y$ and $\vm{\hat \Sigma}_y$ with the Taylor series of second order can be found in \citet{Roth2011efficient}.

\subsubsection{Stirling's interpolation}
The linearization using the Taylor series approximation requires the gradient of the propagation function and thus the analytical calculation of the derivatives of $\vm \func$. However, the derivatives can also be calculated numerically using the central difference quotient with a step size $h$, which results in $2 N_\inp +1$ function evaluations. For one-dimensional normal distributions, it was shown in \citet{Noergaard2000Advances} that the step size $h=\sqrt{3} \text{Var}(\inp)$ is optimal. The mean and the covariance matrix of $\vm y$ can be calculated with the first-order Stirling's interpolation as \citep{Noergaard2000New,Simandl2009Derivative}
\begin{subequations} \label{eq:Stirling_mean_cov_1}
\begin{align}
	\vm{\hat \mu}_y^{(1)} &= \vm \func(\vm \mu_\inp) \label{eq:mean_stirling_interpolation} \\
	\vm{\hat \Sigma}_y^{(1)} &= \frac{1}{4 h^2} \sum_{i = 1}^{N_\inp} \left( \vm \func(\vm \mu_\inp \!+\! h \vm s_i) \!-\! \vm \func(\vm \mu_\inp \!-\! h \vm s_i) \right) \left( \vm \func(\vm \mu_\inp \!+\! h \vm s_i) \!-\! \vm \func(\vm \mu_\inp \!-\! h \vm s_i) \right)\trans \,, \label{eq:covariance_stirling_interpolation}
\end{align}
\end{subequations}
where $\vm s_i$ is the $i$-th column of the Cholesky factor $\vm S$ with
\begin{align}
	\vm \Sigma_\inp = \vm S \vm S\trans \,. \label{eq:cholesky_factorization}
\end{align}
Furthermore, it is possible to approximate the function $\vm \func$ locally by a parabola, which results in the Stirling's interpolation of second order.
In this case, the mean and the covariance matrix of $\vm y$ can be calculated by \citep{Simandl2009Derivative, Zhao2011second, Wu2006Numerical}
\begin{subequations}
\begin{align}%
	\vm{\hat \mu}_y^{(2)} &= \frac{h^2-N_\inp}{h^2} \vm \func(\vm \mu_\inp) + \sum_{i = 1}^{N_\inp} \frac{\vm \func(\vm \mu_\inp + h \vm s_i) + \vm \func(\vm \mu_\inp - h \vm s_i)}{2 h^2} \label{eq:mean_stirling_2}\\
\begin{split}
	\vm{\hat \Sigma}_y^{(2)} &= \vm{\hat \Sigma}_y^{(1)} + \frac{h^2 - 1}{4 h^4} \sum_{i = 1}^{N_\inp} \Big( \left(  \vm \func(\vm \mu_\inp + h \vm s_i) + \vm \func(\vm \mu_\inp - h \vm s_i) - 2 \vm \func(\vm \mu_\inp) \right)\\[-5mm]
	\phantom{\vm{\hat \Sigma}_y^{(2)}} \,&\,\,\phantom{= \vm{\hat \Sigma}_y^{(1)} + \frac{h^2 - 1}{4 h^4} \sum_{i = 1}^{N_\inp}\Big(}\left( \vm \func(\vm \mu_\inp + h \vm s_i) + \vm \func(\vm \mu_\inp - h \vm s_i) - 2 \vm \func(\vm \mu_\inp) \right)\trans \Big) \,,
\end{split}
\label{eq:var_stirling_2}
\end{align}%
\end{subequations}%
where $\vm{\hat \Sigma}_y^{(1)}$ is given by~\eqref{eq:covariance_stirling_interpolation}.

\subsubsection{Unscented transformation}
Similar to Stirling's interpolation, the unscented transformation (UT) also attempts to represent a distribution using $2N_\inp +1 $ sampling points, which in this context are referred to as sigma points \citep{Julier1997New}. These are selected as \citep{Wan2000unscented, Julier2004Unscented}
\begin{subequations} \label{eq:unscented_sigma_points}
\begin{align}
	\vm \inp^{(0)} &= \vm \mu_\inp \\
	\vm \inp^{(i)} &= \vm \mu_\inp +  \sqrt{\alpha^2 (N_\inp + \kappa)} \, \vm s_i \,,\quad i = 1,\, \dots,\, N_\inp \\
	\vm \inp^{(N_\inp+i)} &= \vm \mu_\inp - \sqrt{\alpha^2 (N_\inp + \kappa)} \, \vm s_i \,,\quad i = 1,\, \dots,\, N_\inp \,,
\end{align}
\end{subequations}
where $\alpha$ and $\kappa$ are parameters and $\vm s_i$ is the i-th column of the Cholesky factor $\vm S$ of the covariance matrix $\vm \Sigma_\inp$ as shown in~\eqref{eq:cholesky_factorization}. The parameters $\alpha$ and $\kappa$ can be used to specify the distance of points to the mean of $\vm \inp$. Each of the sigma points is propagated through the function $\vm \func$ on the basis of which the mean and covariance matrix of $\vm y$ can be calculated by a weighted sum. To this end, the weights \citep{Wan2000unscented, Julier2004Unscented}
\begin{subequations} \label{eq:unscented_weights}
\begin{alignat}{3}
	w_\mu^{(0)} &= 1 - \frac{N_\inp}{\alpha^2 (N_\inp + \kappa)}
	\,,\quad &&
	w_\Sigma^{(0)} = w_\mu^{(0)} + (1 - \alpha^2 + \beta)
	\\
	w_\mu^{(i)} &= w_\Sigma^{(i)} = \frac{1}{2 \alpha^2 (N_\inp + \kappa)}
	\,,\quad &&
	i = 1,\, \dots,\, 2 N_\inp
\end{alignat}
\end{subequations}
are introduced, where the weights for the calculation of the mean are denoted with the subscript $\mu$ and the weights for the calculation of the covariance matrix are denoted with the subscript $\Sigma$. The parameter $\beta$ can be used to incorporate prior knowledge about the distribution of $\vm \inp$. For normal distributions, $\beta = 2$ is optimal \citep{Wan2000unscented}. Using the weights of the UT in \eqref{eq:unscented_weights}, the mean and covariance matrix of $\vm y$ is calculated by
\begin{subequations} \label{eq:ut_mean_and_cov}
\begin{align}
	\vm{\hat \mu}_y &= \sum_{i=0}^{2 N_\inp} w_\mu^{(i)} \vm \func(\vm \inp^{(i)}) \label{eq:mean_unscented_transformation}\\
	\vm{\hat \Sigma}_y &= \sum_{i=0}^{2 N_\inp} w_\Sigma^{(i)} \left( \vm \func(\vm \inp^{(i)}) - \vm{\hat \mu}_y \right) \left( \vm \func(\vm \inp^{(i)}) - \vm{\hat \mu}_y \right)\trans \,.\label{eq:covariance_unscented_transformation}
\end{align}
\end{subequations}
An advantage of the UT is that it does not require the gradient of $\vm \func$ and the number of sigma points is relatively small, resulting in a relatively low computation time. Compared to Stirling's interpolation, the UT has more parameters and can therefore be better adjusted to the propagation problem.

\subsubsection{Gaussian quadrature} \label{sec:Gaussian_quadrature}
Mean and covariance matrix of $\vm y$ can be determined by the integrals
\begin{subequations} \label{eq:mean_and_cov_integral}
\begin{align}
	\vm \mu_y &= \ex{\vm \func (\vm \inp)} = \int\limits_{\mathbb{D}} p(\vm \inp) \func(\vm \inp) \,  \text{d} \vm \inp \label{eq:mean_exact_integral}\\
	\vm \Sigma_y &= \mathds E \left[ (\vm \func(\vm \inp) - \vm \mu_y) (\vm \func(\vm \inp) - \vm \mu_y)\trans \right] = \!\!\int\limits_{\mathbb{D}} (\vm \func(\vm \inp) - \vm \mu_y) (\vm \func(\vm \inp) - \vm \mu_y)\trans p(\vm \inp) \, \dd \vm \inp \,, \label{eq:covariance_exact_integral}
\end{align}
\end{subequations}
where $\mathbb{D} \subseteq \R^{N_\inp}$ is the support of the probability density function of $\vm \inp$. These integrals cannot be solved analytically for arbitrary nonlinear functions $\vm \func$ and arbitrary probability density functions $p(\vm \inp)$. However, one way of calculating them approximately is Gaussian quadrature, in which an integral is approximated by a weighted sum \citep{Dunik2020,Lee2009comparative}. This results in estimates for~\eqref{eq:mean_exact_integral} and~\eqref{eq:covariance_exact_integral} in the form
\begin{subequations} \label{eq:mean_and_cov_integral_est}
\begin{align}
	\vm{\hat \mu}_y &= \sum_{i=0}^{N_p-1} w^{(i)} \vm \func(\vm \inp^{(i)}) \label{eq:mean_exact_integral_est}\\
	\vm{\hat \Sigma}_y &= \sum_{i=0}^{N_p-1} w^{(i)} \left( \vm \func(\vm \inp^{(i)}) - \vm{\hat \mu}_y \right) \left( \vm \func(\vm \inp^{(i)}) - \vm{\hat \mu}_y \right)\trans \,, \label{eq:covariance_exact_integral_est}
\end{align}
\end{subequations}
where $N_p= d^{N_\inp}$ is the number of quadrature points, $d$ is the quadrature order, which is a design parameter, $\vm \inp^{(i)}$ is the $i$-th quadrature point, and $w^{(i)}$ is the corresponding weight. The sampling points are the roots of the $d$-th order polynomial from a family of orthogonal polynomials. The choice of polynomials and the associated weights depends on the probability density function $p(\vm \inp)$. GRAMPC-S provides implementations of the Gauss-Hermite quadrature for normal distributions and the Gauss-Legendre quadrature for uniform distributions. However, there are other quadrature rules, which can be found, for example, in \citet{Golub1969Calculation}.

\subsubsection{Monte-Carlo method}
The solution of the integrals~\eqref{eq:mean_and_cov_integral} with Gaussian quadrature is only possible for certain probability density functions, for which the weights and points of the quadrature rule are known. With Monte Carlo simulation, however, these can be approximated for an arbitrary probability density function. For this purpose, $N_p$ samples are randomly selected according to the distribution $p(\vm \inp)$ and these are propagated through the nonlinear function $\vm \func$. In this way, the integrals~\eqref{eq:mean_and_cov_integral} for the stochastic moments can be approximated by \citep{Song2000Monte}%
\begin{subequations} \label{eq:Monte_Carlo_mean_cov}%
\begin{align}
	\vm{\hat \mu}_y &= \frac{1}{N_p} \sum_{i=0}^{N_p-1} \vm \func(\vm \inp^{(i)})\\
	\vm{\hat \Sigma}_y &= \frac{1}{N_p} \sum_{i=0}^{N_p-1} \left( \vm \func(\vm \inp^{(i)}) - \vm{\hat \mu}_y \right) \left( \vm \func(\vm \inp^{(i)}) - \vm{\hat \mu}_y \right)\trans \,.
\end{align}%
\end{subequations}%
For $N_p \rightarrow \infty$, $\vm{\hat \mu}_y$ and $\vm{\hat \Sigma}_y$ converge to $\vm \mu_y$ and $\vm \Sigma_y$, respectively \citep{liu2001monte}. However, the number of sampling points that is required for an accurate estimate of the mean and the covariance matrix is high. This results in a larger computational effort compared to propagation methods with deterministic selection of evaluation points such as UT and Stirling's interpolation.

\subsubsection{Polynomial chaos expansion}
A further widely used method for uncertainty propagation is polynomial chaos expansion (PCE), which approximates the nonlinear function by a sum of weighted polynomial basis functions~\citep{Wiener1938Homogeneous,Paulson2019efficient}. In the following, PCE is first presented for the one-dimensional case $N_\inp = N_y = 1$ and then extended to the multi-dimensional case. To propagate the moments of $y$, it is approximated by 
\begin{align}
	\hat y = \hat \func(\inp) = \sum_{i=0}^{M-1} a_i \phi_i(\inp)\,, \label{eq:pce_onedimensional}
\end{align}
where $\phi_i$ is the $i$-th polynomial from a family of orthogonal polynomials, $a_i$ is the related coefficient and $M$ is the maximum polynomial order. The probability density function $p(\inp)$ determines the choice of basis functions. For example, Hermite polynomials are used for normal distributions and the Legendre polynomials for uniform distributions. An overview of probability density functions and associated basis functions can be found in~\citet{Lucor2004Generalized}. For given basis functions, the mean and variance of $y$ can be approximated with \citep{Paulson2019efficient}
\begin{align}
 \hat \mu_y &= a_0\\
 \hat \sigma_y &= \sum_{i=1}^{M-1} a_i^2 \, \langle \phi_i,\, \phi_i \rangle\,,
\end{align}
which require the coefficients $a_i$. These are given by \citep{Kim2013}
\begin{align}
	a_i = \frac{ \langle \func,\, \phi_i \rangle }{ \langle \phi_i,\, \phi_i \rangle } = \frac{1}{\langle \phi_i,\, \phi_i \rangle}  \int\limits_\mathbb{D} \func(\inp) \phi_i(\inp) w(\inp) \,\dd \inp \,,\quad  i = 0,\dots, M-1 \,, \label{eq:pce_projection}
\end{align}
where $w$ is a weight function. GRAMPC-S solves the integral in \eqref{eq:pce_projection} with Gaussian quadrature as described in Section~\ref{sec:Gaussian_quadrature}. In the multidimensional case, the input $\vm \inp$ and the output $\vm y$ are vectors, requiring the utilisation of multivariate polynomial basis functions $\psi_i$ and vector-valued coefficients $\vm a_i$ with $i \in [0, \, M-1]$, see \citet{Paulson2019efficient} for details.

\subsection{Chance constraint evaluation} \label{sec:chance_constr}

\begin{figure}
	\captionsetup[subfigure]{justification=centering}
	\centering
	\begin{subfigure}[]{0.49\textwidth}
		\centering\hfill
		\pgfmathdeclarefunction{gauss}{2}{%
  \pgfmathparse{1/(#2*sqrt(2*pi))*exp(-((x-#1)^2)/(2*#2^2))}%
}

\begin{tikzpicture}
\begin{axis}[
  no markers, domain=-3.5:2.5, samples=100,
  axis lines=center, xlabel=\small $y$, ylabel=\small $p(y)$, 
  xmin=-5, xmax=3, ymin=-0.03, ymax=0.45,
  axis line style = thick,
  every axis/.append style={font=\small},
  every axis y label/.style={at=(current axis.above origin),anchor=west},
  every axis x label/.style={at=(current axis.right of origin),anchor=north},
  height=5cm, width=6.5cm,
  xticklabel=\empty,
  yticklabel=\empty,
  xtick={0}, ytick=\empty, xtick=\empty,
  ]
  \begin{scope}[on background layer]
  	\addplot [fill=blue1!20, draw=none, domain=-4.5:1.5,forget plot,on layer=axis background] {gauss(-0.5,1)} \closedcycle;
  \end{scope}
  \addplot [blue1, thick] {gauss(-0.5,1)};
  \addplot +[mark=none, thick , black, dashed] coordinates {(-0.5, 0.3989) (-0.5, 0)};
  \addplot +[mark=none, thick , black, dashed] coordinates {(-1.5, 0) (-1.5, 0.242)};
  
  \addplot [black, thick, domain=-1.5:-0.5, samples=2] {-0.02};
  \addplot +[mark=none, thick , black,solid] coordinates {(-1.5, -0.03) (-1.5, -0.01)};
  \addplot +[mark=none, thick , black,solid] coordinates {(-0.5, -0.01) (-0.5, -0.03)};

\end{axis}

\node at (2.4,-0.28) {\small $\sqrt{\text{Var}[y]}$};

\node at (3.24,0.05) {\small $0$};
\node at (4,0.03) {\small $y_\alpha$};
\draw [] (4, 0.26)  -- (4, 0.16);

\node at (4.2,2) {\small $\alpha$};
\draw [] (4.11, 1.88)  -- (3.4, 0.7);

\node at (4.9,1.4) {\small $1 - \alpha$};
\draw [] (4.16, 0.3)  -- (4.85, 1.25);

\node at (1,1.7) {\small $\mathds{E} [y]$};
\draw [] (1.2, 1.5)  -- (2.7, 1);
\end{tikzpicture}
		\subcaption{Chance constraint is not satisfied.}
		\label{plot_constr_approx_1}
	\end{subfigure} \hfill
	\begin{subfigure}[]{0.49\textwidth}
		\centering
		\pgfmathdeclarefunction{gauss}{2}{%
  \pgfmathparse{1/(#2*sqrt(2*pi))*exp(-((x-#1)^2)/(2*#2^2))}%
}

\begin{tikzpicture}
\begin{axis}[
  no markers, domain=-4.9:1, samples=100,
  axis lines=center, xlabel=\small $y$, ylabel=\small $p(y)$, 
  xmin=-5, xmax=3, ymin=-0.03, ymax=0.45,
  axis line style = thick,
  every axis/.append style={font=\small},
  every axis y label/.style={at=(current axis.above origin),anchor=west},
  every axis x label/.style={at=(current axis.right of origin),anchor=north},
  height=5cm, width=6.5cm,
  xticklabel=\empty,
  yticklabel=\empty,
  xtick={-3, -1, 1}, ytick=\empty,xtick=\empty,
  ]
  
  \begin{scope}[on background layer]
  	\addplot [fill=blue1!20, draw=none, domain=-4.9:0,forget plot,on layer=axis background] {gauss(-2,1)} \closedcycle;
  \end{scope}
  \addplot [blue1, thick] {gauss(-2,1)};
  \addplot [black, thick, domain=-2:0.0, samples=2] {-0.02};
  \addplot +[mark=none, thick , black] coordinates {(-0.0, -0.03) (-0.0, -0.01)};
  \addplot +[mark=none, thick , black] coordinates {(-2, -0.03) (-2, -0.01)};
  
  \addplot +[mark=none, thick , black, dashed] coordinates {(-2, 0.3989) (-2, 0)};

\end{axis}

\node at (3.24,0.05) {\small $0$};

\node at (4,2) {\small $\alpha$};
\draw [] (3.86, 1.89)  -- (2.4, 0.7);

\node at (4.7,1.4) {\small $1 - \alpha$};
\draw [] (3.25, 0.3)  -- (4.6, 1.25);

\node at (2.35,-0.28) {\small $z(\alpha) \sqrt{\text{Var}[y]}$};

\end{tikzpicture} \hfill
		\vspace{-0.5mm}
		\subcaption{Chance constraint is satisfied.}
		\label{plot_constr_approx_2}
	\end{subfigure}%
	\caption{Probability density functions of two probabilistic constraints. The shaded area represents the probability of constraint satisfaction.}
	\label{fig_constr}
\end{figure}
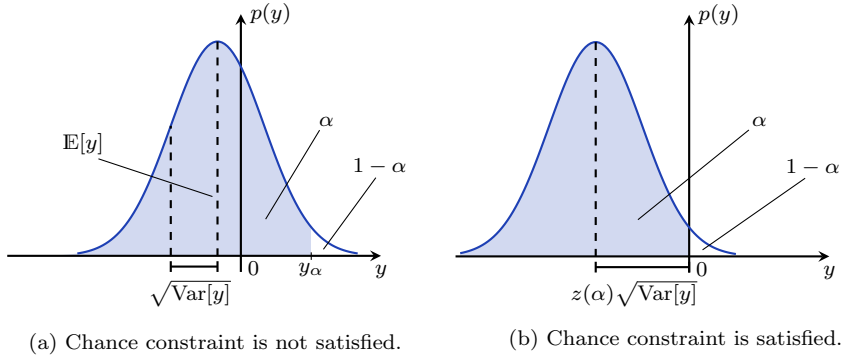

The stochastic OCP \eqref{eq_ocp} should be reformulated as a deterministic OCP. To this end, the chance constraints \eqref{eq_ocp_con} and \eqref{eq_ocp_conTerm} must be evaluated or conservatively estimated. The evaluation of a single constraint $h$ as a function of uncertain inputs can be considered as an uncertainty propagation problem
\begin{align} \label{eq:constr_y}
	y =  h(\vm{x}, \, \vm{u}, \, t) \,,
\end{align}
where the output $y$ is a random variable. The mean and the variance of $y$ can be calculated with any propagation method described in Section~\ref{sec:uncert_prop}. The corresponding chance constraint is 
\begin{align} \label{eq:chance_constr_y}
	\mathds{P} \left[y \leq 0 \right] \geq \alpha \,,
\end{align}
where $\alpha$ is the lower bound of the probability of constraint satisfaction. Figure~\ref{fig_constr} shows an example of the probability density function of the constraint function \eqref{eq:constr_y} for uncertain arguments. The example considers a Gaussian distribution, but in general the distribution can be arbitrary. The probability of constraint satisfaction can be calculated by 
\begin{align}
	\mathds{P}\left[y \leq 0 \right] = \int\limits_{-\infty}^{0} p(y) \,\text{d} y\,.
\end{align}
Thus, the fulfillment of the chance constraint \eqref{eq:chance_constr_y} requires that the $\alpha$-quantile $y_\alpha$, i.e. the solution of the equation
\begin{align} \label{eq:quantile}
 \int\limits_{-\infty}^{y_\alpha} p(y) \,\text{d} y = \alpha	\,,
\end{align}
fulfills $y_\alpha \leq 0$. The area that corresponds to the integral in \eqref{eq:quantile} is shown as a shaded area in Figure~\ref{fig_constr}.

\begin{figure}
	\centering
	\input{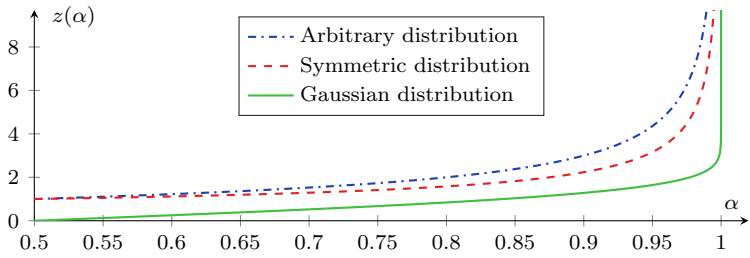}
	\caption{Coefficient $z(\alpha)$ for the constraint tightening depending on the assumption on the probability density function of $y$.}
	\label{fig_constr_coeff}
\end{figure}

The calculation of $y_\alpha$ in \eqref{eq:quantile} is generally not possible as it requires knowledge of $p(y)$. As described in Section~\ref{sec:uncert_prop}, the exact propagation of uncertainties for nonlinear propagation functions is challenging, which is why most propagation methods focus on the calculation of stochastic moments. As shown in Figure~\ref{plot_constr_approx_2}, satisfaction of the chance constraint for distributions with finite variance can also be described by the tightened inequality constraint
\begin{align}
\Ex \left[y\right] \leq z(\alpha) \sqrt{\text{Var} \left[y\right]} \label{eq:tightened_constraint}
\end{align}
as a function of the mean and variance of $y$, where $z(\alpha)$ is a coefficient that depends on assumptions on the distribution. If only the mean and variance of $y$ are known but no other characteristics of the distribution, the coefficient can be derived from Chebyshev's inequality as \citep{Calafiore2006, Marshall1960}
\begin{align}
z(\alpha) = \sqrt{\frac{\alpha}{1-\alpha}} \,.
\end{align}
Since Chebyshev's inequality applies to any distribution with finite variance, it is usually conservative. The conservatism of the estimate can be reduced if additional properties of the distribution are known. If it is known that the distribution of $y$ is symmetrical, the coefficient can be chosen to
\begin{align}
 z(\alpha) = \sqrt{\frac{1}{2 \left( 1-\alpha\right)}} \,.
\end{align}
A third option is available if it can be assumed that $p(y)$ is a Gaussian distribution, where $z(\alpha)$ can be calculated as the $\alpha$-quantile of the standard normal distribution, which can be precalculated. Figure~\ref{fig_constr_coeff} shows the function $z(\alpha)$ for the three assumptions on the distribution of $y$ that are presented above. It can be seen that the assumption on the distribution has a significant influence on the coefficient, in particular for large values of $\alpha$. Further options for the choice of $z(\alpha)$ would be conceivable if the type of distribution is known.

\subsection{Deterministic reformulation of the stochastic optimal control problem} \label{sec:det_reformulation}

\begin{figure}
	\centering
	\begin{tikzpicture}[thick]
		\node[draw, align=center] (stoch) at (0,0) {Stochastic OCP \\[2mm] \( \small \begin{aligned}
				\min_{\vm u} \;\;& \mathds E \left[ V(\vm x(T), \vm p) \right] + \int_0^T \mathds E \left[ l(\vm x, \vm u, \vm p) \right] \,\dd t \\
				\operatorname{s.\!t.} \;\;& \dd \vm x = \vm f(\vm x, \vm u, \vm p) \,\dd t + \vm \sigma_w \,\dd \vm w \,,\;\; \vm x(0) = \vm x_0 \\
				& \mathds P \left[ h_i(\vm x, \vm u) \leq 0 \right] \geq \alpha_i \,,\;\; i = 1, \dots, N_h \\
				& \vm u \in [\vm u_{\min}, \vm u_{\max}]
			\end{aligned} \)};
		
		\node[draw, align=center] (sbr) at (-3.3,-2.5) {Sampling-based representation \\[2mm] \( \small \begin{aligned}
				\vm{\tilde x} &= \begin{bmatrix} \vm x^{(1)} & \dots & \vm x^{(N_s)} \end{bmatrix} &&\!\!\!\in \mathds R^{N_x \times N_s}\\
				\vm{\tilde p} &= \begin{bmatrix} \vm p^{(1)} & \dots & \vm p^{(N_s)} \end{bmatrix} &&\!\!\!\in \mathds R^{N_p \times N_s}
			\end{aligned} \)};
		
		\node[draw, align=center] (mbr) at (3.3,-2.5) {Moment-based representation \\[2mm] \( \small \begin{aligned}
		\vm{\tilde x} &= \begin{bmatrix} \vm \mu_x & \vm \Sigma_x & \vm \Sigma_{xp} \end{bmatrix} &&\!\!\!\in \mathds R^{N_x \times (1 + N_x + N_p)} \\
		\vm{\tilde p} &= \begin{bmatrix} \vm \mu_p & \vm \Sigma_p \end{bmatrix} &&\!\!\!\in \mathds R^{N_p \times (1 + N_p)}
	\end{aligned} \)};
		
		\node[draw, align=center] (det) at (0,-5) {Deterministic  approximation \\[2mm] \( \small \begin{aligned}
				\min_{\vm u} \quad& \tilde V(\vm{\tilde x}(T), \vm{\tilde p}) + \int_0^T \tilde l(\vm{\tilde x}, \vm u, \vm{\tilde p}) \,\dd t \\
				\operatorname{s.\!t.} \quad& \vm{\dot{\tilde x}} = \vm{\tilde f}(\vm{\tilde x}, \vm u, \vm{\tilde p})  \,,\quad \vm{\tilde x}(0) = \vm{\tilde x}_0 \\
				& \tilde h_i(\vm{\tilde x}, \vm u) \leq 0 \,,\quad i = 1, \dots, N_h \\
				& \vm u \in [\vm u_{\min}, \vm u_{\max}]
		\end{aligned} \)};
	
		\draw[->] (stoch) -| (sbr);
		\draw[->] (stoch) -| (mbr);
		\draw[->] (sbr) |- (det);
		\draw[->] (mbr) |- (det);
	\end{tikzpicture}
	\caption{Deterministic approximation of the stochastic OCP either by sampling-based or moment-based representation of probability distributions.}
	\label{fig:approximation_overview}
\end{figure}
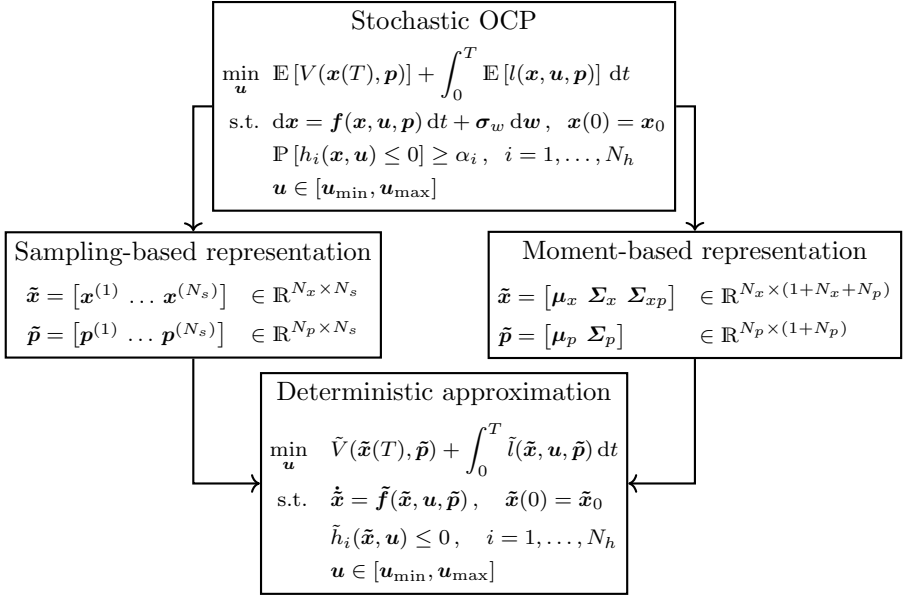

The previously described methods are used to derive a deterministic approximation of the stochastic optimal control problem \eqref{eq_ocp} that can be handled by existing solvers as, in our case, the GRAMPC toolbox. To this end, the probability density functions of the random variables must be represented by a finite number of variables. As Figure~\ref{fig:approximation_overview} shows, GRAMPC-S provides two types of representation: Sampling-based representation (SR) and moment-based representation (MR). In the first case, the states \( \vm x(t) \), \( t \in [0, T] \) and the parameters \( \vm p\) are approximated by \( N_s \) sampling points. In the second case, the means \( \vm \mu_x(t) \approx \mathds E[\vm x(t)] \), \( \vm \mu_p = \mathds E[\vm p] \) and covariance matrices \( \vm \Sigma_x(t) \approx \cov[\vm x(t)] \), \( \vm \Sigma_{xp}(t) \approx \cov[\vm x(t), \vm p] \), \( \vm \Sigma_p = \cov[\vm p] \) are utilized to represent the random variables. For reasons of readability, only the deterministic reformulation of inequality constraints \eqref{eq_ocp_con} is discussed in this section. However, terminal inequality constraints \eqref{eq_ocp_conTerm} can be reformulated in the same way and GRAMPC-S provides implementations for both types of constraints.

\subsubsection{Sampling-based representation}

This first approach approximates the probability distributions of the initial state \( \vm x_0 \) and the parameters \( \vm p \) by \( N_s \) sampling points \( \vm x^{(i)} \in \mathds R^{N_x} \) and \( \vm p^{(i)} \in \mathds R^{N_p} \) with \( i = 1, \dots, N_s \).
The samples can be chosen either in a randomized (Monte Carlo) or deterministic (e.g.\ unscented transformation) manner.
The deterministic state and parameter variables are therefore defined as
\begin{equation}
	\vm{\tilde x} = \begin{bmatrix} \vm x^{(1)} & \dots & \vm x^{(N_s)} \end{bmatrix} \in \mathds R^{N_x \times N_s}
	\,,\quad
	\vm{\tilde p} = \begin{bmatrix} \vm p^{(1)} & \dots & \vm p^{(N_s)} \end{bmatrix} \in \mathds R^{N_p \times N_s} \,.
\end{equation}
Without the Wiener process \( \vm w(t) \), the dynamics~\eqref{eq_ocp_dyn} can be formulated independently per sample as
\begin{equation}
	\vm{\dot x}^{(i)} = \vm f(\vm x^{(i)}, \vm u, \vm p^{(i)}) \,,\quad \vm x^{(i)}(0) = \vm x_0^{(i)} \,,\quad i = 1, \dots, N_s\,,
	\label{eq:dyn_sampling_based}
\end{equation}
which can be summarized as
\begin{equation}
	\underbrace{\begin{bmatrix} \vm{\dot x}^{(1)} & \dots & \vm{\dot x}^{(N_s)} \end{bmatrix}}_{\vm{\dot{\tilde x}}} = \underbrace{\begin{bmatrix} \vm f(\vm x^{(1)}, \vm u, \vm p^{(1)}) & \dots & \vm f(\vm x^{(N_s)}, \vm u, \vm p^{(N_s)}) \end{bmatrix}}_{\vm{\tilde f}(\vm{\tilde x}, \vm u, \vm{\tilde p})} \,.
\end{equation}
Considering a Wiener process is more difficult.
Using a sufficient number of random samples, as in the Monte Carlo approach, the diffusion could be applied to each sample separately, e.g.\ by performing an Euler-Maruyama integration.
However, this would require modifications of the underlying solvers of GRAMPC which perform the integration.
Using the unscented transformation as example of a method with deterministic sampling points, the Wiener process could be considered by the square root prediction \citep[Algorithm 4.5]{Sarkka2007}.
Due to the added complexity, GRAMPC-S currently supports the Wiener process only in the moment-based representation. 

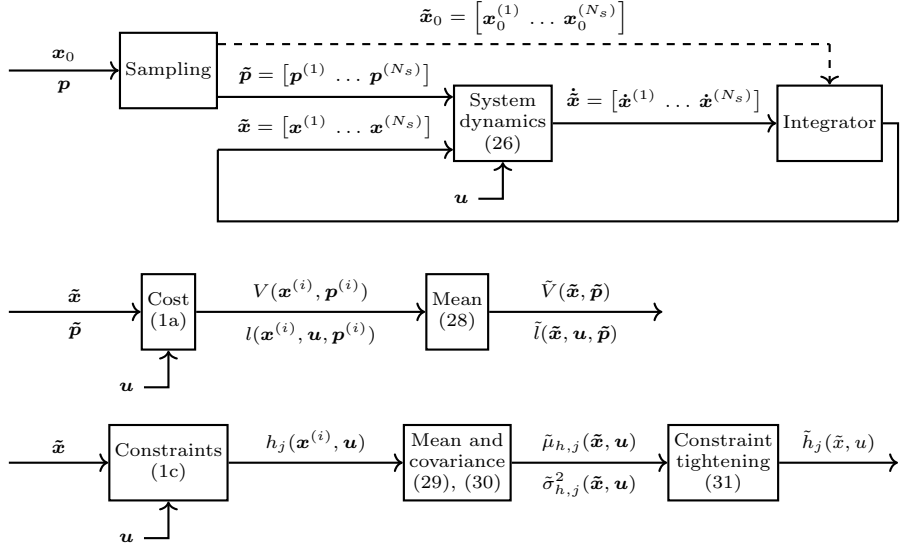
\begin{figure}
	\centering
	\begin{tikzpicture}[thick,font=\scriptsize]
		\node[draw, align=center, minimum height=1cm, inner sep=0.7mm] (sampling) at (-5.3, 2) {Sampling};
		\node[draw, align=center, minimum height=1cm, inner sep=0.7mm] (dyn) at (-0.9, 1.3) {System\\dynamics\\\eqref{eq:dyn_sampling_based}};
		\node[draw, align=center, minimum height=1cm, inner sep=0.7mm] (integrator) at (3.4, 1.3) {Integrator};
		\node[align=center] (u) at (-1.45, 0.3) {$\vm u$};
		\coordinate (p) at (-7.4, 2);
		
		\coordinate (rp1) at (4.3, 0);
		\coordinate (lp1) at (-4.65, 0.95);
		\coordinate (p1) at (3.4, 2.35);
		\coordinate (p11) at (-1.5, 2.15);
		
		\draw[->] (p) to node[above ,midway] {$\vm x_0$} node[below, midway] {$\vm p$}(sampling);
		\draw[dashed] ([yshift=0.35cm]sampling.east) -- node[above, midway] {$\vm{\tilde x}_0 = \begin{bmatrix} \vm x_0^{(1)} & \dots & \vm x_0^{(N_s)} \end{bmatrix}$} (p1);
		\draw[->] ([yshift=-0.35cm]sampling.east) -- node[above, midway] {$\vm{\tilde p} = \begin{bmatrix} \vm p^{(1)} & \dots & \vm p^{(N_s)} \end{bmatrix}$} ([yshift=0.35cm]dyn.west);
		\draw[dashed, ->] (p1) -- (integrator);
		\draw[->] (dyn) -- node[above, midway] {$\vm{\dot{\tilde x}} = \begin{bmatrix} \vm{\dot x}^{(1)} & \dots & \vm{\dot x}^{(N_s)} \end{bmatrix}$} (integrator);
		\draw (integrator) -| (rp1);
		\draw (rp1) -| (lp1);
		\draw[->] (lp1) -- node[above, midway] {$\vm{\tilde x} = \begin{bmatrix} \vm x^{(1)} & \dots & \vm x^{(N_s)} \end{bmatrix}$} ([yshift=-0.35cm]dyn.west);
		\draw[->] (u) -| (dyn);

		\node[draw, align=center, minimum height=1cm, inner sep=0.7mm] (cost) at (-5.3, -1.2) {Cost\\\eqref{eq_ocp_cost}};
		\node[draw, align=center, minimum height=1cm, inner sep=0.7mm] (meanAndCov2) at (-1.5, -1.2) {Mean\\\eqref{eq:mean_cost_samples}};
		\node[align=center] (u2) at (-5.85, -2.2) {$\vm u$};
		\coordinate (p2) at (-7.4, -1.2);
		\coordinate (out) at (1.2, -1.2);
		
		\draw[->] (u2) -| (cost);
		\draw[->] (p2) to node[below, midway] {$\vm{\tilde p}$} node[above, midway] {$\vm{\tilde x}$} (cost.west);
		\draw[->] (cost) to node[midway, above] {$V(\vm x^{(i)},\vm p^{(i)})$} node[midway, below, align=center] {$l(\vm x^{(i)},\vm u,\vm p^{(i)})$}(meanAndCov2);
		\draw[->] (meanAndCov2) to node[midway, above, align=center] {$\tilde V(\vm{\tilde x}, \vm{\tilde p})$} node[midway, below, align=left] {$\tilde l(\vm{\tilde x}, \vm u, \vm{\tilde p})$} (out);

		\node[draw, align=center, minimum height=1cm, inner sep=0.7mm] (constraint) at (-5.3, -3.2) {Constraints\\\eqref{eq_ocp_con}};
		\node[draw, align=center, minimum height=1cm, inner sep=0.7mm] (meanAndCov3) at (-1.5, -3.2) {Mean and\\covariance\\\eqref{eq:mean_hj}, \eqref{eq:cov_hj}};
		\node[align=center] (u3) at (-5.85, -4.2) {$\vm u$};
		\coordinate (p3) at (-7.4, -3.2);
		\node[align=center, draw, minimum height=1cm, inner sep=0.7mm] (constrT) at (2, -3.2) {Constraint\\tightening\\\eqref{eq:deterministic_constr}};
		\coordinate (out2) at (4.3, -3.2);
		
		\draw[->] (u3) -| (constraint);
		\draw[->] (p3) to node[above, midway] {$\vm{\tilde x}$} (constraint.west);
		\draw[->] (constraint) to node[midway, above, align=center] {$h_j(\vm x^{(i)},\vm u)$}(meanAndCov3);
		\draw[->] (meanAndCov3) to node[midway, above, align=center] {$\tilde \mu_{h,j}(\vm{\tilde x}, \vm u)$} node[midway, below, align=left] {$\tilde \sigma_{h,j}^2(\vm{\tilde x}, \vm u)$} (constrT);
		\draw[->] (constrT) -- node[above, midway] {$\tilde h_j (\tilde x, u)$} (out2);
	\end{tikzpicture}
	
	\caption{Prediction of the deterministic states $\vm{\tilde x} = \begin{bmatrix} \vm x^{(1)} & \dots & \vm x^{(N_s)} \end{bmatrix}$ using sampling-based representation (top), evaluation of cost functions (middle), and evaluation of the tightened constraints (bottom).}
	\label{fig:sampling_overview}
\end{figure}

The expected values in the cost functional~\eqref{eq_ocp_cost} are approximated by the weighted averages
\begin{equation}
	\mathds E[V(\vm x, \vm p)] \approx \underbrace{\sum_{i=1}^{N_s} w_{\mu}^{(i)} V(\vm x^{(i)}, \vm p^{(i)})}_{\tilde V(\vm{\tilde x}, \vm{\tilde p})}
	\,,\quad
	\mathds E[l(\vm x, \vm u, \vm p)] \approx \underbrace{\sum_{i=1}^{N_s} w_{\mu}^{(i)} l(\vm x^{(i)}, \vm u, \vm p^{(i)})}_{\tilde l(\vm{\tilde x}, \vm u, \vm{\tilde p})}
	\label{eq:mean_cost_samples}
\end{equation}
where \( w_{\mu}^{(i)} \), \( i = 1, \dots, N_s \) are the weighting factors for computing mean values, which depend on the uncertainty propagation method.
In order to approximate the inequality constraints, their mean
\begin{equation}
	\mathds E[h_j(\vm x, \vm u)] \approx \tilde \mu_{h,j}(\vm{\tilde x}, \vm u) = \sum_{i=1}^{N_s} w_{\mu}^{(i)} h_j(\vm x^{(i)}, \vm u) \label{eq:mean_hj}
\end{equation}
and covariance
\begin{equation}
	\cov[h_j(\vm x, \vm u)] \approx \tilde \sigma_{h,j}^2(\vm{\tilde x}, \vm u) = \sum_{i=1}^{N_s} w_{\sigma}^{(i)} \left( h_j(\vm x^{(i)}, \vm u) - \tilde \mu_{h,j}(\vm{\tilde x}, \vm u) \right)^2 \label{eq:cov_hj}
\end{equation}
can be estimated from the sampling points, where \( w_{\sigma}^{(i)} \), \( i = 1, \dots, N_s \) are the weighting factors for computing covariances.
The deterministic approximation of the constraints is then
\begin{equation}
	\underbrace{\tilde \mu_{h,j}(\vm{\tilde x}, \vm u) + z_{j} \left( \alpha_j \right) \sqrt{\tilde \sigma_{h,j}^2(\vm{\tilde x}, \vm u)}}_{\tilde h_j(\vm{\tilde x}, \vm u)} \leq 0 \label{eq:deterministic_constr}
\end{equation}
with the constraint tightening coefficient \( z_{j} \left( \alpha_j \right) \) depending on the desired probability threshold \( \alpha_j \). Figure~\ref{fig:sampling_overview} presents an overview of the state prediction, cost evaluation, and constraint tightening for the sampling-based representation.

In case of the Monte Carlo approach, the constraint can alternatively be enforced individually per sample, that is
\begin{equation}
	\underbrace{h_j(\vm x^{(i)}, \vm u) \leq 0 \,,\quad i = 1, \dots, N_s}_{\vm{\tilde h}_j(\vm{\tilde x}, \vm u) \leq \vm 0} \,.
\end{equation}
The more random samples are used the higher is the confidence that the chance constraint is actually satisfied with a probability of \( \alpha_j \).
Consider a set of \( N_s \) samples, then the sample proportion \( \tilde \alpha_j \) is approximately normal distributed with 
\begin{equation}
	\tilde \alpha_j \sim \mathcal N\left( \alpha_j, \frac{\alpha_j (1 - \alpha_j)}{N_s} \right)
\end{equation}
if \( N_s \alpha_j \geq 10\) and \( N_s (1 - \alpha_j) \geq 10 \).
For example, if \( N_s = 100 \) samples satisfy the constraint, one can state that the chance constraint is satisfied with a probability \( \alpha_j = 0.95 \) with a confidence of \( 0.99 \), while the confidence for the probability \( \alpha_j = 0.99 \) is only \( 0.84 \).
This allows to relate the number of samples \( N_s \), the desired probability \( \alpha_j \), and the confidence to each other.

\subsubsection{Moment-based representation} \label{sec:MR}

The second approach approximates the probability distributions of the state \( \vm x \) and the parameters \( \vm p \) by their mean values \( \vm \mu_{x} = \mathds E[\vm x] \), \( \vm \mu_p = \mathds E[\vm p] \), their covariances \( \vm \Sigma_{x} = \cov[\vm x] \), \( \vm \Sigma_p = \cov[\vm p] \), and their cross-covariance \( \vm \Sigma_{xp} = \cov[\vm x, \vm p] \).
The deterministic state and parameter variables are therefore defined as
\begin{equation}
	\vm{\tilde x} = \begin{bmatrix} \vm \mu_x & \vm \Sigma_x & \vm \Sigma_{xp} \end{bmatrix} \in \mathds R^{N_x \times (1 + N_x + N_p)}\,, \quad
		\vm{\tilde p} = \begin{bmatrix} \vm \mu_p & \vm \Sigma_p \end{bmatrix} \in \mathds R^{N_p \times (1 + N_p)} \,.
\end{equation}
In view of the stochastic differential equation, the means and covariances evolve over time according to the ordinary differential equations \citep{Sarkka2007}
\begin{equation}
	\begin{aligned}
		\vm{\dot \mu}_x &= \frac{\dd}{\dd t} \mathds E[\vm x] = \mathds E[\vm f(\vm x, \vm u, \vm p)] \\
		\vm{\dot \Sigma}_x &= \frac{\dd}{\dd t} \cov[\vm x] = \cov[\vm f(\vm x, \vm u, \vm p), \vm x] + \cov[\vm f(\vm x, \vm u, \vm p), \vm x]\trans + \vm \Sigma_w \\
		\vm{\dot \Sigma}_{xp} &= \frac{\dd}{\dd t} \cov[\vm x, \vm p] = \cov[\vm f(\vm x, \vm u, \vm p), \vm p]
	\end{aligned}
	\label{eq:mean_cov_dynamics}
\end{equation}
with the covariance matrix \( \vm \Sigma_w = \vm \sigma_w \vm \sigma_w\trans \) depending on the diffusion of the Wiener process. Any of the introduced propagation methods can be applied to approximate the expected value \( \mathds E[\vm f(\vm x, \vm u, \vm p)] \) and the covariance matrices \( \cov[\vm f(\vm x, \vm u, \vm p), \vm x] \) and \( \cov[\vm f(\vm x, \vm u, \vm p), \vm p] \).
Similarly, the expected values of the costs \( \mathds E[\vm V(\vm x, \vm p)] \approx \tilde V(\vm{\tilde x}, \vm{\tilde p}) \), \( \mathds E[l(\vm x, \vm u, \vm p)] \approx \tilde l(\vm{\tilde x}, \vm u, \vm{\tilde p}) \) as well as the mean and covariance of the constraints \( \mathds E[h_j(\vm x, \vm u)] \), \( \cov[h_j(\vm x, \vm u)] \) can be approximated. Based on the mean and covariance of the constraints, the deterministic approximation of the constraint is calculated using~\eqref{eq:deterministic_constr}.

GRAMPC-S distinguishes between two moment-based representation approaches, which differ in the utilized uncertainty propagation method. The first approach uses a first-order Taylor expansion while the second approach is based on sampling-based propagation methods. Using the first-order Taylor expansion as propagation method, the mean values are computed as
\begin{subequations}
	\begin{align}
		\mathds E[\vm f(\vm x, \vm u, \vm p)] &\approx \vm f(\vm \mu_x, \vm u, \vm \mu_p) \label{eq:mean1}\\
		\mathds E[V(\vm x, \vm p)] &\approx \tilde V(\vm{\tilde x}, \vm{\tilde p}) = V(\vm \mu_x, \vm \mu_p) \label{eq:mean2}\\
		\mathds E[l(\vm x, \vm u, \vm p)] &\approx \tilde l(\vm{\tilde x}, \vm u, \vm{\tilde p}) = l(\vm \mu_x, \vm u, \vm \mu_p) \label{eq:mean3}\\
		\mathds E[h_j(\vm x, \vm u)] &\approx \tilde \mu_{h,j}(\vm{\tilde x}, \vm u) = h_j(\vm \mu_x, \vm u) \label{eq:mean4}
	\end{align}
\end{subequations}
while the covariances are computed as
\begin{subequations}
	\begin{align}
		\cov[\vm f(\vm x, \vm u, \vm p), \vm x] &\approx \frac{\partial \vm f(\vm \mu_x, \vm u, \vm \mu_p)}{\partial \vm x} \vm \Sigma_x + \frac{\partial \vm f(\vm \mu_x, \vm u, \vm \mu_p)}{\partial \vm p} \vm \Sigma_{xp} \label{eq:cov1}\\
		\cov[\vm f(\vm x, \vm u, \vm p), \vm p] &\approx \frac{\partial \vm f(\vm \mu_x, \vm u, \vm \mu_p)}{\partial \vm x} \vm \Sigma_{xp} + \frac{\partial \vm f(\vm \mu_x, \vm u, \vm \mu_p)}{\partial \vm p} \vm \Sigma_{p} \label{eq:cov2}\\
		\cov[h_j(\vm x, \vm u)] &\approx \tilde \sigma_{h,j}^2(\vm{\tilde x}, \vm u) = \frac{\partial h_j(\vm \mu_x, \vm u)}{\partial \vm x} \vm \Sigma_x \frac{\partial h_j(\vm \mu_x, \vm u)}{\partial \vm x}\trans\,. \label{eq:cov3}
	\end{align}
\end{subequations}
The dynamics \eqref{eq:mean_cov_dynamics} are fully defined by \eqref{eq:mean1}, \eqref{eq:cov1}, and \eqref{eq:cov2}. The constraints of the deterministic OCP are given by \eqref{eq:deterministic_constr} using the means \eqref{eq:mean4} and the variances \eqref{eq:cov3}.

\begin{figure}
	\centering
	\begin{tikzpicture}[thick,font=\scriptsize]
		\node[draw, align=center, inner sep=0.7mm, minimum height=1cm] (integrator) at (-2.25,0.1) {Integrator};
		\node[draw, align=center, minimum height=1cm, inner sep=0.7mm] (sampling) at (-4.5, 2) {Sampling};
		\node[draw, align=center, minimum height=1cm, inner sep=0.7mm] (dyn) at (-0.7, 2) {System\\dynamics\\\eqref{eq:dyn_sampling_based}};
		\node[draw, align=center, minimum height=1cm, inner sep=0.7mm] (meanAndCov) at (2.7, 2) {Mean and\\covariance};
		\node[draw, align=center, minimum height=1cm, inner sep=0.7mm] (momentDyn) at (2.7, 0.1) {Mean and\\covariance\\dynamics \eqref{eq:mean_cov_dynamics}};
		\node[align=center] (u) at (-1.25, 1) {$\vm u$};
		\coordinate (p) at (-7.4, 2.35);
		
		\coordinate (rp1) at (4.1, 2);
		\coordinate (integ1) at (-2.25,-1);
		\coordinate (rp2) at (4.1, 0.1);
		\coordinate (lp1) at (-5.9, 1.65);
		\coordinate (lp2) at (-5.9, 0.1);
		\coordinate (p1) at (0.45, 2.75);
		\coordinate (c) at (-1.8, 2);
		\node[circle,fill=black,inner sep=0pt,minimum size=3pt] (circ) at (c) {};
		
		\draw[->] (sampling.east) -- node [pos=0.4,above] {$\vm x^{(i)},\, \vm p^{(i)}$} (dyn.west);
		\draw[->] ([yshift=-0.35cm]dyn.east) to node [midway,above] {$\vm f(\vm x^{(i)}\!, \vm u, \vm p^{(i)})$} ([yshift=-0.35cm]meanAndCov.west);
		\draw (meanAndCov) -- (rp1);
		\draw (rp1) -- (rp2);
		\draw (integrator) to node [midway,above] {$\vm{\tilde x} = \begin{bmatrix} \vm \mu_x & \vm \Sigma_x & \vm \Sigma_{xp} \end{bmatrix}$} (lp2);
		\draw (lp2) -- (lp1);
		\draw[->] (lp1) -- ([yshift=-0.35cm]sampling.west); 
		\draw[->] (rp2) to (momentDyn);
		\draw[->] (momentDyn) to node [midway,above] {$\vm{\dot{\tilde x}} = \begin{bmatrix} \vm{\dot \mu}_x & \vm{\dot \Sigma}_x & \vm{\dot \Sigma}_{xp} \end{bmatrix}$} (integrator);
		\draw[dashed, ->] (1.45, -1) -- node[above, midway] {$\vm{\tilde x}_0 = \begin{bmatrix} \vm \mu_{x_0} & \vm \Sigma_{x_0} & \vm \Sigma_{x_0p} \end{bmatrix}$} (integ1) -- (integrator);
		\draw[->] (u) -| (dyn);
		\draw (c) |- (p1);
		\draw[->] (p1) |- ([yshift=0.35cm]meanAndCov.west);
		\draw[->] (p) -- node[above, midway] {$\vm{\tilde p} = \begin{bmatrix} \vm \mu_p & \vm \Sigma_p \end{bmatrix}$} ([yshift=0.35cm]sampling.west);

		\node[draw, align=center, minimum height=1cm, inner sep=0.7mm] (sampling2) at (-6.1, -2) {Sampling};
		\node[draw, align=center, minimum height=1cm, inner sep=0.7mm] (cost) at (-3.8, -2) {Cost\\\eqref{eq_ocp_cost}};
		\node[draw, align=center, minimum height=1cm, inner sep=0.7mm] (meanAndCov2) at (-0.7, -2) {Mean\\\eqref{eq:mean_cost_samples}};
		\node[align=center] (u2) at (-4.4, -3) {$\vm u$};
		\coordinate (p2) at (-7.4, -2);
		\coordinate (out) at (1.55, -2);
		
		\draw[->] (u2) -| (cost);
		\draw[->] (p2) to node[below, midway] {$\vm{\tilde p}$} node[above, midway] {$\vm{\tilde x}$} (sampling2.west);
		\draw[->] (sampling2.east) to node [midway,above] {$\vm x^{(i)}$} node [midway,below] {$\vm p^{(i)}$} (cost.west);
		\draw[->] (cost) to node[midway, above] {$V(\vm x^{(i)},\vm p^{(i)})$} node[midway, below, align=center] {$l(\vm x^{(i)},\vm u,\vm p^{(i)})$}(meanAndCov2);
		\draw[->] (meanAndCov2) to node[midway, above, align=center] {$\tilde V(\vm{\tilde x}, \vm{\tilde p})$} node[midway, below, align=left] {$\tilde l(\vm{\tilde x}, \vm u, \vm{\tilde p})$} (out);

		\node[draw, align=center, minimum height=1cm, inner sep=0.7mm] (sampling3) at (-6.1, -4) {Sampling};
		\node[draw, align=center, minimum height=1cm, inner sep=0.7mm] (constraint) at (-3.8, -4) {Constraints\\\eqref{eq_ocp_con}};
		\node[draw, align=center, minimum height=1cm, inner sep=0.7mm] (meanAndCov3) at (-0.7, -4) {Mean and\\covariance\\\eqref{eq:mean_hj}, \eqref{eq:cov_hj}};
		\node[align=center] (u3) at (-4.4, -5) {$\vm u$};
		\coordinate (p3) at (-7.4, -4);
		\node[align=center, draw, minimum height=1cm, inner sep=0.7mm] (constrT) at (2.3, -4) {Constraint\\tightening\\\eqref{eq:deterministic_constr}};
		\coordinate (out2) at (4.4, -4);
		
		\draw[->] (u3) -| (constraint);
		\draw[->] (p3) to node[above, midway] {$\vm{\tilde x}$} (sampling3.west);
		\draw[->] (sampling3.east) to node [midway,above] {$\vm x^{(i)}$} (constraint.west);
		\draw[->] (constraint) to node[midway, above, align=center] {$h_j(\vm x^{(i)},\vm u)$}(meanAndCov3);
		\draw[->] (meanAndCov3) to node[midway, above, align=center] {$\tilde \mu_{h,j}(\vm{\tilde x}, \vm u)$} node[midway, below, align=left] {$\tilde \sigma_{h,j}^2(\vm{\tilde x}, \vm u)$} (constrT);
		\draw[->] (constrT) -- node[above, midway] {$\tilde h_j (\tilde x, u)$} (out2);
	\end{tikzpicture}
	 \vspace{-2mm}
	\caption{Prediction of the deterministic states $\vm{\tilde x} = \begin{bmatrix} \vm \mu_x & \vm \Sigma_x & \vm \Sigma_{xp} \end{bmatrix}$ using moment-based representation with sampling-based uncertainty propagation methods (top), evaluation of cost functions (middle), and evaluation of the tightened constraints (bottom). Each function evaluation requires the representation of  $\vm{\tilde x}$ and $\vm{\tilde p}$ by $N_s$ sampling points $\vm x^{(i)}$ and $\vm p^{(i)}$ with $i = 1, \dots, N_s$.}
	\label{fig:resampling}
\end{figure}
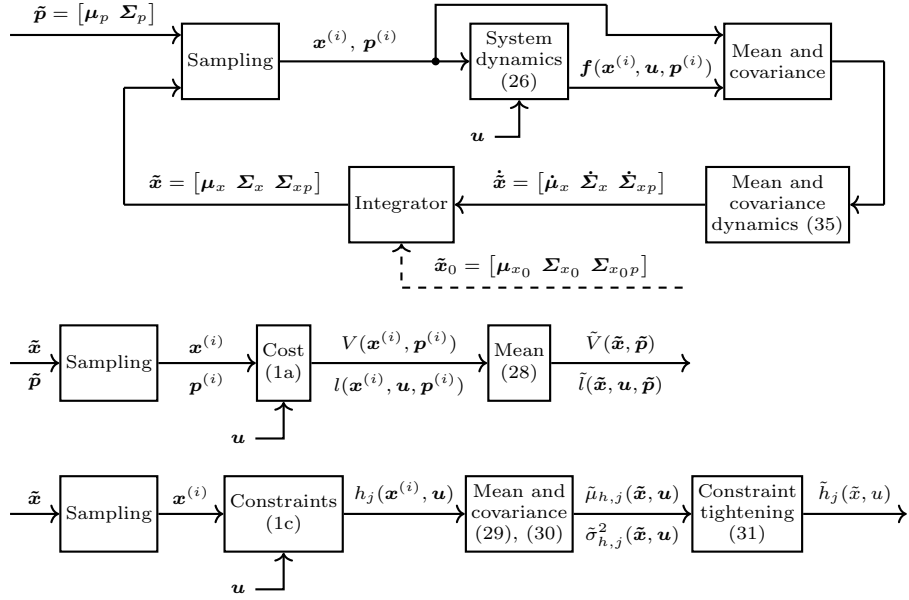

The second moment-based approach also represents the uncertain states by $\vm{\tilde x} = \begin{bmatrix} \vm \mu_x & \vm \Sigma_x & \vm \Sigma_{xp} \end{bmatrix}$, but does not use a linearization of the system dynamics. Instead, sampling-based methods are used whenever uncertainties must be propagated through potentially nonlinear functions. Figure~\ref{fig:resampling} presents an overview of the state prediction (top), cost evaluation (middle), and evaluation of the tightened constraint (bottom). In order to calculate the time derivative of the deterministic states, $N_s$ sampling points $\vm x^{(i)}$ and $\vm p^{(i)}$ with $i = 1, \dots , N_s$ are generated and propagated through the system dynamics \eqref{eq:dyn_sampling_based}. Based on the initial sampling points and the propagated ones, the expected value $\mathds E[\vm f(\vm x, \vm u, \vm p)]$ and the cross-covariances $\cov[\vm f(\vm x, \vm u, \vm p), \vm x]$ and $\cov[\vm f(\vm x, \vm u, \vm p), \vm p]$ are calculated. These are utilized to evaluate the time derivative of the deterministic state using~\eqref{eq:mean_cov_dynamics}. The deterministic cost and constraint approximations also require the generation of sampling points. The deterministic cost approximations $\tilde V(\vm{\tilde x}, \vm{\tilde p})$ and $\tilde l(\vm{\tilde x}, \vm u, \vm{\tilde p})$ are calculated by the weighted average \eqref{eq:mean_cost_samples} of the costs of the sampling points. Similarly, the mean and covariance of the constraints are calculated using \eqref{eq:mean_hj} and \eqref{eq:cov_hj}. These are used to formulate the tightened constraints $\tilde h_j (\tilde x, u)$ for $i = 1,\dots, N_h$ using~\eqref{eq:deterministic_constr}.

\subsection{Consideration of Gaussian processes} \label{sec:GP}
The application of model predictive control requires a model of the system dynamics. This model is typically a simplification of the actual behavior and includes errors that can degrade the performance of the control. In order to avoid this loss of performance, a more accurate model would be necessary, but its identification can be time-consuming. An alternative to complicated models is the combination of a simple model, which can be identified easily, and Gaussian process regression (GPR). To this end, the unknown part of the system dynamics is represented by a Gaussian process (GP) and data points are collected that represent the unmodeled parts of the system dynamics \citep{Hewing.2020}. The mean and the variance of the GP can then be evaluated by means of the data points. Since the variance of the GP is larger than zero in most cases, the consideration of a GP in the system dynamics results in uncertain predicted states.

In the following, the system dynamics \eqref{eq_ocp_dyn} is extended to%
\begin{align}%
\text{d}\vm{x} = \vm f(\vm{x},\, \vm u,\, \vm p,\, t)\, \text{d}t +  \vm g \left( \vm x,\, \vm u \right)\text{d}t + \vm \sigma_w \, \text{d} \vm w \,,\label{eq:SDE_with_GP}%
\end{align}%
where $\vm f$ represents the known part of the system dynamics and $\vm g$ represents the unknown part. The function $\vm g$ is not accessible and thus, it is approximated by a multivariate GP 
\begin{align}
\vm d \left( \vm z \right) \sim \mathcal{N}\left( \vm \mu_d \left( \vm z \right),\, \vm \Sigma_d \left( \vm z \right) \right)
\end{align}
with $\vm z = [\vm x\trans,\, \vm u\trans]\trans$, mean $\vm \mu_d \left( \vm z \right) = [\mu_d^1,\, \mu_d^2,\, ...,\, \mu_d^{N_x}]\trans$, and covariance matrix $\vm \Sigma_d \left( \vm z \right) = \text{diag}( [\sigma_d^1,\, \sigma_d^2,\,...,\, \sigma_d^{N_x}]\trans )$. The elements of $\vm d = [d_1, \, \dots, \, d_{N_x}]\trans$ can be regarded as independent univariate GPs with their individual kernels $k^i(\cdot, \, \cdot)$, $i \in [1, \, N_x]$. The calculation of the mean and variance of these GPs is based on $M$ input data points $\vm z_j^\text{in} = [\vm x_j\trans,\, \vm u_j\trans]\trans$ with $j \in [1, \, M]$ and corresponding output data points $\vm z_j^\text{out} = \vm g(\vm x_j, \vm u_j) + \vm \nu_j$ with measurement noise $\vm \nu \sim \mathcal{N}(0,\, \text{diag}(\vm \sigma_\nu)) \in \R^{N_x}$, where $\vm \sigma_\nu$ is the vector of measurement noise variances. Using $\vm z^\text{out} = [z_1^{\text{out}}, \,...,\, z_M^{\text{out}}]\trans \in \R^{M \times N_x}$, the mean and variance of the $i$-th univariate GP can be calculated by \citep{Rasmussen2006}
\begin{align}
\mu_d^i (\vm z) &= \Ex \left[  d_i (\vm z)\right] = \vm k_*^{i\,}(\vm z)\trans \left( \vm K^i + \sigma_{\nu, i} \vm I \right)^{-1} \left[\vm z^\text{out}\right]_i \label{eq:GP_mean}\\
\sigma_d^i (\vm z) &= \text{Var}\left[ d_i (\vm z) \right] = k^i_{**}(\vm z) - \vm k_*^{i\,}(\vm z)\trans \left( \vm K^i + \sigma_{\nu, i} \vm I \right)^{-1} \vm k^i_*(\vm z)\,, \label{eq:GP_var}
\end{align}
where $\vm I$ is the identity matrix, $\left[\vm z^\text{out}\right]_i$ is the $i$-th column of $\vm z^\text{out}$, $k_*^i(\vm z)$ is a $M$-dimensional vector whose $j$-th element is $k_{*j}^i = k^i(\vm z,\, \vm z_j^\text{in})$, $\vm K^i$ is a $M \times M$-dimensional matrix whose $jj'$-th element is $\vm K^i_{jj'} = k^i(\vm z_j^\text{in},\, \vm z_{j'}^\text{in})$, and $k^i_{**}(\vm z) = k^i(\vm z,\, \vm z)$ is a scalar.

Considering the variance of the Gaussian process in the sampling-based representation is difficult for the same reasons why the Wiener process is difficult to consider.
Therefore, GRAMPC-S currently supports Gaussian processes only in the moment-based representation as described in Section~\ref{sec:MR} by adapting the calculation of $\vm{\dot \mu}_x$, $\vm{\dot \Sigma}_x$, and $\vm{\dot \Sigma}_{xp}$.
Using $\vm y = \vm f(\vm{x},\, \vm u,\, \vm p,\, t) + \vm \mu_d \left( \vm z \right)$, the ordinary differential equations \eqref{eq:mean_cov_dynamics} are modified to
\begin{align}
	\vm{\dot \mu}_x &= \frac{\dd}{\dd t} \mathds E[\vm x] = \mathds E[\vm y] \\
	\vm{\dot \Sigma}_x &= \frac{\dd}{\dd t} \cov[\vm x] = \cov[\vm y, \vm x] + \cov[\vm y, \vm x]\trans + \vm \Sigma_d(\vm \mu_z) + \vm \Sigma_w \\
	\vm{\dot \Sigma}_{xp} &= \frac{\dd}{\dd t} \cov[\vm x, \vm p] = \cov[\vm y, \vm p]
\end{align}
with $\vm \mu_z = \Ex [\vm z] = [\vm \mu_x\trans,\, \vm u\trans]\trans$. 
The main difference is that the variance \( \vm \Sigma_d \) of the GP is evaluated for the state mean \( \vm \mu_x \) and taken into account in the dynamics of the state covariance \( \vm \Sigma_x \).
The calculation of the deterministic cost and constraint approximation is not affected by the consideration of GPs in the system dynamics.
\section{Structure of GRAMPC-S} \label{sec:structure}

\begin{figure}[b]
	\centering
	\begin{tikzpicture}
	\tikzset{font=\small}

 \draw[align=center] (0,0)  node (a) {GRAMPC-S};
 \draw[align=center] (2.5,1)  node (b) {Point\\Transformation} edge [-] (a);
 \draw[align=center] (0,-1)  node (c) {Constraint\\Approximation} edge [-] (a);
 \draw[align=center] (2.5,-1)  node (d) {Problem\\Description} edge [-] (a);
 \draw[align=center] (-2.5,1)  node (e) {Distribution} edge [-] (a);
 \draw[align=center] (0,1)  node (f) {Simulator} edge [-] (a);
 \draw[align=center] (-2.5,-1) node (g) {Gaussian\\Process} edge [-] (a);
 \draw[align=center] (2.5,0)  node (h) {Stationary\\Kernel} edge [-] (a);
 \draw[align=center] (-2.5,0)  node (i) {Orthogonal\\Polynomials} edge [-] (a);
 
 \draw[align=center] (-1.5,2)  node (k) {Gaussian\\Distribution} edge [-] (e);
 \draw[align=center] (-3.5,2)  node (l) {Uniform\\Distribution} edge [-] (e);
 \draw[align=center] (-3.4,1.35)  node {...};  
 
 \draw[align=center] (1.5,2)  node (n) {Unscented\\Transformation} edge [-] (b);
 \draw[align=center] (3.5,2)  node (o) {Monte-Carlo\\Simulation} edge [-] (b);
 \draw[align=center] (3.5,1.4)  node {...};  
 
 \draw[align=center] (-1.7,-2)  node (p) {Gaussian} edge [-] (c);
 \draw[align=center] (0,-2.03)  node (q) {Symmetric} edge [-] (c);
 \draw[align=center] (1.7,-2.03)  node (r) {Arbitrary} edge [-] (c);
 
 \draw[align=center] (4.7,-1.4)  node (s) {Sampling\\Points} edge [-] (d);
 \draw[align=center] (3.6,-2)  node (t) {Linearization} edge [-] (d);
 \draw[align=center] (2.7,-1.64)  node  {...};
 
 \draw[align=center] (-5,0.5)  node (u) {Hermite\\Polynomials} edge [-] (i);
 \draw[align=center] (-5,-0.5)  node (v) {Legendre\\Polynomials} edge [-] (i); 
 
 \draw[align=center] (5,0.5)  node (w) {Squared\\Exponential} edge [-] (h);
 \draw[align=center] (5,-0.5)  node (x) {Locally\\Periodic} edge [-] (h);
 \draw[align=center] (3.75,-0.6)  node[rotate=90]  {...};
 
\end{tikzpicture} \vspace{-1mm}
	\caption{General structure of GRAMPC-S with interfaces and implementations.}
	\label{fig:GRAMPC_S_structure}
\end{figure}
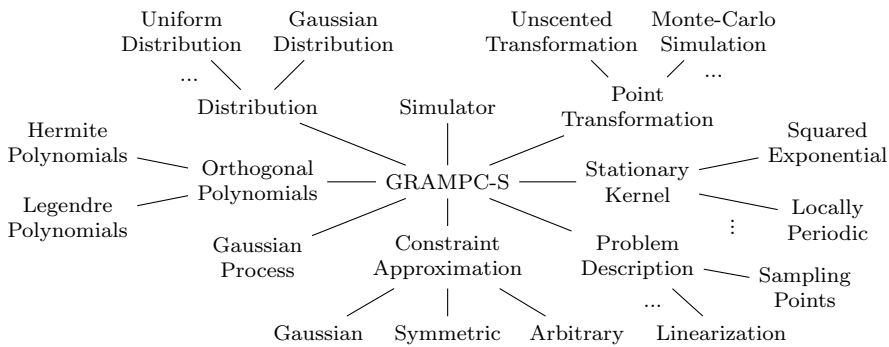

The code of GRAMPC-S is written in C++ and the only required dependency is the Eigen library for the implementation of linear algebra\footnote{Guennebaud G, Jacob B, et al. (2010) Eigen v3. http://eigen.tuxfamily.org}. The underlying solver GRAMPC is based on the augmented Lagrangian formulation and uses a gradient method to solve the inner optimization problem. Therefore, GRAMPC-S provides not only the deterministic reformulation of the cost function, the system dynamics and the constraints of the OCP, but also the respective gradients of these functions. Figure~\ref{fig:GRAMPC_S_structure} shows the general structure of GRAMPC-S.

The stochastic representation is based on distributions for which the mean and variance can be evaluated. In addition, the Monte Carlo simulations necessitate the generation of random samples according to the respective probability density functions. A list of all implemented distributions can be found in the GRAMPC-S manual. These distributions can be passed to a point transformation, which represents them by a finite number of sampling points and calculates the stochastic moments of the underlying distribution. Uncertainty propagation using Gaussian quadrature and PCE also requires the generation of orthogonal polynomials that correspond to the respective distributions.

The actual approximation of a stochastic OCP by a deterministic OCP takes place in the problem description, whereby a distinction is made between sampling-based representation, moment-based representation with linearization and moment-based representation using sampling points. To formulate the OCP, the choice of constraint approximation must be specified, where the three variants from Section~\ref{sec:chance_constr} are available. The approach using moment-based representation allows to take a Gaussian process for approximating an unknown part of the system dynamics into account as described in Section~\ref{sec:GP}. In addition to the input and output data of the Gaussian process, the corresponding kernel with hyperparameters must be specified. The list of available kernels can be found in the manual of GRAMPC-S. Once the problem description has been defined, it can be passed to the solver. The system can then be simulated and the results can be plotted, e.g. in Matlab. Moreover, GRAMPC-S provides an S-function for use in Simulink. In addition to the simulation in Simulink, this allows to test the controller on experimental setups equipped, for example, with dSPACE hardware.
\section{Evaluation} \label{sec:evaluation}
This section evaluates the performance of GRAMPC-S and the related computation times. Section~\ref{sec:cstr} focuses on the example of a chemical reactor with uncertain system parameters, for which both open-loop prediction and closed-loop control are examined. Subsequently, in Section~\ref{sec:nonlin_chain}, the dependence of the computation time on the number of states is analysed using the example of a nonlinear chain. Section~\ref{sec:exp_val_DI_GP} presents the usage of Gaussian processes in system dynamics. In Section~\ref{sec:exp_val_inv_pendulum}, the application of GRAMPC-S is experimentally validated for the control of an inverted pendulum.

\subsection{Control of a continuous stirred tank reactor} \label{sec:cstr}
An illustrative example of a stochastic OCP is a simplified model of a continuous stirred tank reactor with the state $\vm x = [c_A\,, c_B]\trans$, where $c_A$ is the normalized concentration of an educt and $c_B$ is the normalized concentration of a product. The normalized dynamics of the chemical process are given by \citep{Perez.2002,Graichen2004}
\begin{subequations} \label{eq_reactor_dyn}
\begin{align}
	\dot c_A &= -p_1 c_A - p_3 c_A^2 + (1 - c_A) u\\
	\dot c_B &= p_1 c_A - p_2 c_B - c_B u
\end{align}
\end{subequations}
with the normalized flow rate as control input $u$. The system parameters $\vm p = [p_1, \, p_2,\, p_3]\trans$ in \eqref{eq_reactor_dyn} are considered as uncertain with the distributions
\begin{align}
 \hspace{-0.1cm}p_1 \sim \mathcal{U}\left(48,\, 52\right)\,\si{\per\hour}\,, \quad p_2 \sim \mathcal{U}\left(95,\, 105\right)\,\si{\per\hour}\,, \quad p_3 \sim \mathcal{U}\left(95,\, 105\right)\,\si{\per\hour}\,. \label{eq:cstr_param}
\end{align}
The objective of the control is to reach and stabilize the desired state $\vm x_\text{des}$ and the desired control input $u_\text{des}$ while satisfying a chance constraint for $c_b$. The considered OCP is 
\begin{subequations} \label{eq:cstr_problem}
\begin{align}
 \min_{u} \,\,\quad& J\left( u; \, \vm x_0 \right) = \mathds{E} \left[\int\limits_0^T \!
\Delta \vm x \trans \begin{bmatrix}
	 10^2 & 0 \\ 0 & 10^2
\end{bmatrix} \Delta \vm x  + 0.1 \;\!\Delta u^2 \,\, \text{d}t \right] \label{eq_cost_reactor}\\
 \operatorname{s.t.} \quad\;\,& \dot{\vm{x}} = \vm f(\vm{x}, \, u, \, \vm p)\,,\quad \vm{x}(0) = \vm x_0 \label{dyn_ocp_reactor} \\
 & \mathds{P} \left[ c_B \leq 0.14 \right] \geq 0.9 \label{eq_prob_constr_reactor}\\
 & u \in \left[ 10 ,\, 100 \right]
\end{align}%
\end{subequations}%
with $\Delta \vm x = \vm x - \vm x_\text{des}$ and $\Delta u = u - u_\text{des}$. In addition to the three uncertain system parameters, a measurement noise for the two states with variance $\sigma_x^2$ is taken into account, with the result that the system has five uncertain variables. The stochastic OCP~\eqref{eq:cstr_problem} is solved with GRAMPC-S and the deterministic reformulations of the stochastic OCP from Section~\ref{sec:det_reformulation} as well as the uncertainty propagation methods from Section~\ref{sec:uncert_prop} are compared, in particular with regard to the required computation time.

\begin{figure}
	\centering
	\input{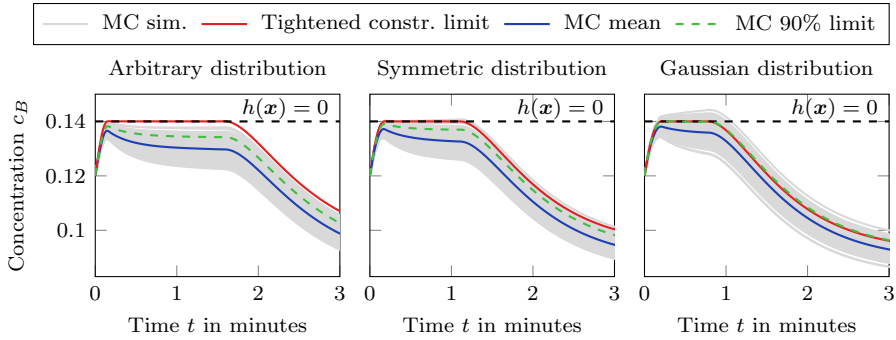}
	\caption{Concentration $c_B$ in 1000 Monte-Carlo simulations (gray) for three different assumptions on the probability density function of the constraint. The blue and the green lines show the mean and the 90\% quantile of $c_B$ according to the Monte-Carlo simulations. The red lines show the boundaries of the tightened constraint \eqref{eq:tightened_constraint}.}
	\label{fig_openloop_cstr}
\end{figure}

In a first step, the open-loop prediction of the state trajectories is investigated for three assumptions on the probability density function of the constraints presented in Section~\ref{sec:chance_constr}. For this purpose, the OCP \eqref{eq:cstr_problem} is solved using GRAMPC-S with sampling-based representation and unscented transformation. Figure~\ref{fig_openloop_cstr} shows the results of the concentration $c_B$ for 1000 Monte-Carlo simulations representing the uncertain parameters \eqref{eq:cstr_param} and the uncertain initial state with $\sigma_x^2 = 10^{-9}$ using the control input calculated by the controller. The blue and the green lines show the mean and the 90\% quantile of $c_B$ according to the Monte-Carlo simulations and the red lines show the boundaries of the tightened constraints \eqref{eq:tightened_constraint} based on the assumption on the probability density function of the constraint.

In all three cases, the concentration $c_B$ initially increases until the boundary of the tightened constraint is reached. Thereafter, $c_B$ remains at this boundary for some time until it decreases. A comparison of the three plots shows that the 90\% quantile of $c_B$ according to the Monte-Carlo simulations is significantly below the boundary $c_B = 0.14$ if an arbitrary or a symmetric distribution is assumed. In this example, these assumptions thus lead to an overly cautious behavior of the controlled system. In contrast, the assumption of a Gaussian distribution causes the 90\% quantile of $c_B$ to be almost exactly on the boundary $c_B = 0.14$, thus avoiding an excessive tightening of the chance constraint \eqref{eq_prob_constr_reactor}.  

\begin{figure}
	\centering
	\input{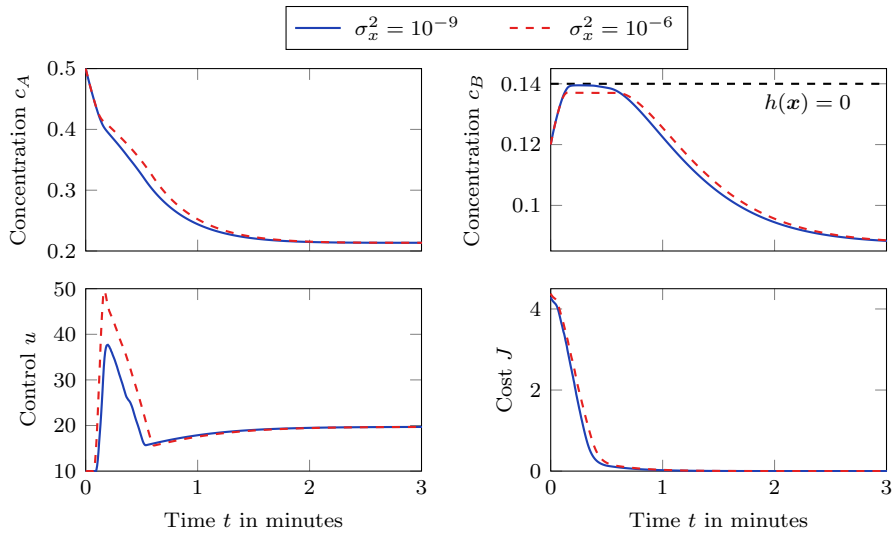}
	\caption{State and input trajectories and cost for the continuous stirred tank reactor problem \eqref{eq:cstr_problem} with sampling-based representation and unscented transformation. The inequality constraint is tightened using Chebyshev's inequality.}
	\label{fig_closedloop_cstr}
\end{figure}

In a second step, the trajectories of the states and the input resulting from the closed-loop control of the system \eqref{eq:cstr_problem} with SMPC are investigated. In the considered scenario, the prediction horizon is chosen to $T=\SI{36}{\second}$, the sampling time is $\Delta t = \SI{1}{\second}$, the number of interpolation points of the prediction is set to 20 and the underlying solver GRAMPC is executed with 2 augmented Lagrangian iterations and 2 gradient iterations. The inequality constraint is tightened using Chebyshev's inequality. Figure \ref{fig_closedloop_cstr} shows the resulting state and input trajectories and the cost $J$ for two noise levels $\sigma_x$ computed with the sampling-based representation and unscented transformation. The case with $\sigma_x = 10^{-9}$ corresponds to the same measurement noise as in the prediction in Figure~\ref{fig_openloop_cstr}. However, a comparison shows that in the closed-loop case, the state $c_B$ is considerably closer to the constraint than in the prediction in Figure~\ref{fig_openloop_cstr}. The reason for this is that the uncertainty of the predicted state increases with the prediction time due to the uncertain system parameters. As the time progresses, a new measurement of the state is available in the closed-loop case, which reduces the uncertainty of the state resulting in a smaller safety margin. The red lines in Figure~\ref{fig_closedloop_cstr} correspond to a simulation with $\sigma_x = 10^{-6}$. Due to the higher measurement noise, the uncertainty of the predicted states is larger. Accordingly, the controller operates more cautiously and keeps a larger distance to the constraint of $c_B$.

Of particular interest for SMPC is the computation time, because it limits the applicability in practice. Table~\ref{tab:reactor_compTime} shows the required average computation time per sampling step of GRAMPC-S with the proposed methods. The time $t_\text{CPU}$ is evaluated on a notebook with Windows 10, Intel(R) Core(TM) i5-10210U CPU and a MinGW-w64 compiler. The time $t_\text{dSPACE}$ is evaluated on a dSPACE MicroLabBox with a $\SI{2}{\giga\hertz}$ NXP QorlQ P5020 CPU. The evaluation shows that the computation times with SR are generally lower than for the corresponding counterparts with MR. The reason for this is that with SR the random variables only have to be approximated by points once per MPC sampling step, while with MR the approximation is performed for each evaluation of the cost function, the system dynamics and the constraint. The differences in the computation time of UT and Stirling's interpolation are small because the number of sampling points is the same for these methods.  The computation times of Gaussian quadrature, Monte-Carlo simulation and PCE are higher in comparison, but these methods can achieve a higher approximation accuracy, as shown in \citet{Landgraf.2023}.
\begin{table}
	\setlength{\tabcolsep}{5pt}
	\small
	\renewcommand{\arraystretch}{1.3}
	\caption{Average computation time per sampling step and cost difference of SMPC with sampling-based representation (SR) and moment-based representation (MR) for the continuous stirred tank reactor. The cost reference is generated with Monte-Carlo sampling and $10^6$ sampling points.}
	\centering
	\begin{tabular}{lccccc}
		\toprule
		\multirow{2}{*}{Propagation method} & \multirow{2}{*}{Representation} &\multirow{2}{*}{Parameters}& $t_\text{CPU}$& $t_\text{dSPACE}$ & \multirow{2}{*}{$\left\vert\Delta J\right\vert$}\tabularnewline 
		&&& in ms & in ms&\tabularnewline
		\midrule 
		Unscented transformation & SR && 0.16 & 0.57 & 0.06\tabularnewline
		1st order Stirling's interp. &SR && 0.16 & 0.55 & 0.07\tabularnewline
		2nd order Stirling's interp. &SR && 0.16 & 0.56 & 0.06\tabularnewline
		Gaussian quadrature & SR  &$d=3$& 2.70& 10.20 & 0.01\tabularnewline
		Monte-Carlo sampling & SR &$N_p = 1000$& 11.10 & 45.14 & 0.03\tabularnewline
		PCE & SR & $d=3$, $M=2$& 2.86 & 11.3 & 0.01\tabularnewline
		\midrule
		Taylor linearization& MR && 0.40 & 1.59 & 0.26\tabularnewline
		Unscented transformation & MR && 1.71 & 8.93 & 0.20\tabularnewline
		1st order Stirling's interp. & MR && 1.68 & 8.80 & 0.25\tabularnewline
		2nd order Stirling's interp. & MR && 1.69 & 9.01 & 0.21\tabularnewline
		\bottomrule 
	\end{tabular}
	\label{tab:reactor_compTime}
\end{table}

Besides the computation time, the propagation method influences the performance of the control, wherefore the costs \eqref{eq_cost_reactor} of the trajectories resulting from the closed-loop control are evaluated. The cost of a stochastic model predictive controller with SR and Monte-Carlo sampling with $10^6$ points are used as a reference. Deviations from this reference indicate that the uncertainties of the predicted states are over- or underestimated by the controller. The last column in Table~\ref{tab:reactor_compTime} presents the absolute value of the cost difference $\left\vert\Delta J\right\vert$ between the resulting costs with the respective controller and the reference. It can be seen that methods based on SR result in smaller deviations from the reference than methods based on MR. The cause of the higher errors in MR is likely the continuous approximation of the probability density functions by mean and covariance.  A comparison of the SR methods shows that Gaussian quadrature, Monte-Carlo sampling and PCE achieve relatively low cost differences, because in addition to the mean and variance of the input variables, higher moments of the distributions are also taken into account. However, this is associated with a comparatively high computation time.

\subsection{Control of a nonlinear chain} \label{sec:nonlin_chain}

The second case study considers the motion control of a nonlinear chain with $n+1$ connected elements from \citet{Kirches2012}, which is a common benchmark problem for nonlinear MPC that is used e.g. in \citet{Englert.2019} and \citet{Verschueren.2021}. The first chain element is fixed while the others can move in the three-dimensional space. The state of the middle chain elements is defined by their respective positions and velocities. The control input is the velocity of the final chain element, which results in a total number of states $N_x = 6 n - 3$. A detailed description of the system dynamics is given in \citet{Kirches2012}.

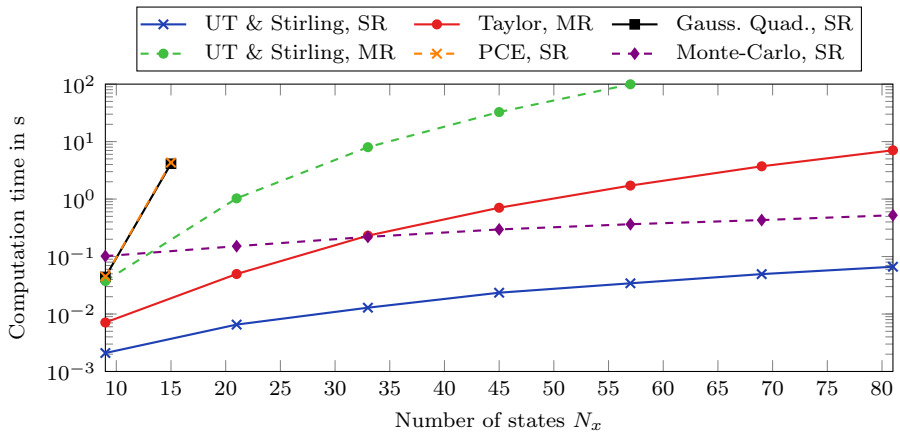
\begin{figure}[b]
	\centering
	\begin{tikzpicture}
\tikzset{font=\small}

\begin{axis}[%
width=4.08in,
height=1.5in,
scale only axis,
xmin=9,
xmax=81,
ymin=1e-3,
ymax=100,
ymode=log,
yminorticks=true,
xlabel={Number of states $N_x$},
ylabel={Computation time in s},
ylabel near ticks,
yminorticks=true,
max space between ticks=20, 
legend cell align={left},
legend columns=3,
legend style = {at={(0.5, 1.27)}, anchor=north, inner sep=1pt, style={column sep=0.16cm}},
]
\addplot [color=blue1, thick, mark=x, mark options={solid, blue1,thick,mark size=2.5pt}]
  table[row sep=crcr]{%
9	2.09546e-3\\
21	6.54418e-3\\
33	12.8789e-3\\
45	23.479e-3\\
57	34.2337e-3\\
69	49.4744e-3\\
81	66.4327e-3\\
};
\addlegendentry{UT \& Stirling, SR}

\addplot [color=red1, thick, mark=*, mark options={solid, red1,thick,mark size=1.5pt}]
  table[row sep=crcr]{%
9	7.12214e-3\\
21	49.6674e-3\\
33	231.001e-3\\
45	706.492e-3\\
57	1717.43e-3\\
69	3717.67e-3\\
81	7058.3e-3\\
};
\addlegendentry{Taylor, MR}

\addplot [color=black, thick, mark=square*, mark options={solid, black,thick,mark size=1.6pt}]
  table[row sep=crcr]{%
9	44.1672e-3\\
15	4121.37e-3\\
};
\addlegendentry{Gauss. Quad., SR}

\addplot [color=green1, dashed, thick, mark=*, mark options={solid, green1,thick,mark size=1.5pt}]
  table[row sep=crcr]{%
9	37.4201e-3\\
21	1028.45e-3\\
33	7997.8e-3\\
45	32662.8e-3\\
57	99371e-3\\
};
\addlegendentry{UT \& Stirling, MR}

\addplot [color=orange, thick, mark=x,dashed, mark options={solid, orange,thick,mark size=2.5pt}]
  table[row sep=crcr]{%
9	44.9558e-3\\
15	4315.38e-3\\
};
\addlegendentry{PCE, SR}

\addplot [color=violet,dashed, thick, mark=diamond*, mark options={solid, violet,thick,mark size=1.8pt}]
  table[row sep=crcr]{%
9	10.1762e-2\\
21	15.1384e-2\\
33	21.9875e-2 \\
45	29.6409e-2\\
57	36.5206e-2\\
69	43.1068e-2\\
81	52.356e-2\\
};
\addlegendentry{Monte-Carlo, SR}

\end{axis}
\end{tikzpicture}%
	\caption{Average computation time per sampling step of a stochastic model predictive controller for the nonlinear chain problem with varying number of chain elements. The parameters of the propagation methods are the same as in Table~\ref{tab:reactor_compTime}. Some methods exceed the available computing resources with increasing numbers of states.}
	\label{fig_chain}
\end{figure}

A stochastic model predictive controller is designed to stabilize the chain based on noisy measurements of the states. The objective of the study is to examine the dependence of computation time of the controller on the number of states. To this end, the number of chain elements is varied from 2 to 14. Figure~\ref{fig_chain} shows the resulting computation time per MPC step for sampling-based and moment-based representation with the available uncertainty propagation methods using the same computer hardware as in Section~\ref{sec:cstr}. The results of UT and Stirling's interpolation are approximately the same and thus the corresponding lines are merged for the sake of clarity. It can be seen from Figure~\ref{fig_chain} that the computation time of SMPC with SR and UT or Stirling's interpolation is the lowest of all methods regardless of the number of states. The control with the same propagation methods require significantly more computation time if MR is used because the sampling points are calculated more often. The computation time of the control with SR and Gaussian quadrature scales poorly because the number of sampling points increases exponentially with the number of states. Accordingly, the evaluation was only possible for the cases $n=2$ and $n=3$, as the computing resources of the system are not sufficient for higher numbers of chain elements. Since the implementation of PCE is based on Gaussian quadrature, the results are correspondingly close to each other. The computation time for the control with Monte Carlo sampling increases relatively slowly because the number of sampling points was chosen to $N_p = 1000$ regardless of the number of chain elements. In practice, however, it can be assumed that $N_p$ must be increased with an increasing number of states in order to represent the probability density functions with sufficient accuracy.

\subsection{Control of a water tank with learned dynamics} \label{sec:exp_val_DI_GP}

\begin{figure}[b]
	\centering
	\begin{tikzpicture}[domain=0:2]
	\tikzset{font=\small}
	
	\draw[thick] (0,0) -- (0,-2) --(2.4,-2);
	\draw[thick] (2.4,-1.9) -- (2,-1.9) -- (2,0);
	
	\draw[thick] (-1, 0.2) -- (0.5, 0.2) -- (0.5, -0.1);
	\draw[thick] (-1, 0.4) -- (0.7, 0.4) -- (0.7, -0.1);
	
	\draw[color=black, samples=100]	plot (\x,{0.02*sin(30*\x r)-1});
	
	\draw[align=center] (-0.2,-1) -- (-0.2, -2) node[midway,left] {$h$};
	\draw[] (-0.13, -1) -- (-0.27, -1);
	\draw[] (-0.13, -2) -- (-0.27, -2);
	
	\draw[align=center] (0,-2.2) -- (2, -2.2) node[midway,below] {$A$};
	\draw[] (0, -2.13) -- (0, -2.27);
	\draw[] (2, -2.13) -- (2, -2.27);
	
	\draw[align=center] (-0.4,-2) -- (-0.4, -1.9) node[midway,left] {$a$};
	\draw[] (-0.47, -2) -- (-0.33, -2);
	\draw[] (-0.47, -1.9) -- (-0.33, -1.9);
	
	\draw[->] (0.6, -0.2) -- (0.6, -0.8) node[midway, right] {$q$};
	\draw[->] (2.5, -1.95) -- (3.3, -1.95) node[midway,above] {$g(h)$};
 
\end{tikzpicture}
	\caption{Schematic representation of the water tank with states and parameters.}
	\label{fig_tank}
\end{figure}
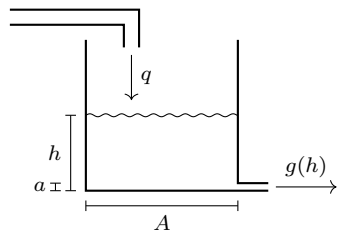

In this section, a water tank is controlled with SMPC, where parts of the system dynamics are learned from noisy data using Gaussian process regression. Figure~\ref{fig_tank} shows a schematic of the water tank with an inflow with volume flow rate $q$, the fill level of the water tank $h$, and an outflow with volume flow rate $g(h)$. The temporal change of $q$ can be controlled by the input $u$ and the volume flow rate of the outflow is given by Torricelli's law. Thus, the dynamics of this system is
\begin{subequations} \label{eq:water_tank_dyn}
\begin{align}
	\dot q &= u\\
	\dot h &= \frac{1}{A} q + g(h)
\end{align}
\end{subequations}
with
\begin{align}
	g(h) = - \frac{a}{A} \sqrt{2 g h}\,,
\end{align}
where $A = \SI{1}{\meter\squared}$ is the cross-sectional area of the water tank, $a=\SI[quotient-mode=fraction,parse-numbers=false]{\frac{1}{30}}{\meter\squared}$ is the cross-sectional area of the drain, and $g = \SI{9.81}{\meter\per\second\squared}$ is the gravitational acceleration. The function $g(h)$ is used for the simulation but is assumed to be unknown for the controller. Instead, the controller uses Gaussian process regression to approximate the unknown function $g(h)$, as described in Section~\ref{sec:GP}. The Gaussian process $d(h)$ is learned from a set of noisy data points using a squared exponential kernel. Figure~\ref{fig_gp_points} shows the volume flow rate $g(h)$ and the mean of the learned function $d(h)$, where the variance of the measurement noise is $\sigma_\nu^2 = \SI[print-unity-mantissa=false]{e{-3}}{\metre\squared\per\second\squared}$. 

\begin{figure}
	\centering
	\input{gp_points.tex}
	\caption{Unknown part of the system dynamics $g(h)$ and Gaussian process $d(h)$, which is learned from data points with measurement noise of variance $\sigma_\nu^2 = \SI[print-unity-mantissa=false]{e{-3}}{\metre\squared\per\second\squared}$.}
	\label{fig_gp_points}
\end{figure}

\begin{figure}[b]
	\centering
	\input{water_tank_results.tex}
	\caption{State and input trajectories of the water tank \eqref{eq:water_tank_dyn} for closed-loop control with SMPC based on a Gaussian process with 10 data points. The data points are generated with a measurement noise variance $\sigma_\nu^2 = \SI[print-unity-mantissa=false]{e{-3}}{\metre\squared\per\second\squared}$ (blue) and $\sigma_\nu^2 = \SI[print-unity-mantissa=false]{e{-9}}{\metre\squared\per\second\squared}$ (red).}
	\label{fig_gp_res}
\end{figure}

The objective of the control is to fill the water tank without exceeding the maximum fill level of $h_\text{max} = \SI{1}{\meter}$. To this end, the cost function of the optimization problem is defined as 
\begin{align}
	J\left( u; \, \vm x_0 \right) = \mathds{E} \left[\int\limits_0^T  \left(h-1\right)^2  +  u^2 \,\, \text{d}t \right]
\end{align}
and the constraints are defined as
\begin{align}
	&u \in \left[\SI{0}{\cubic\meter\per\second\squared},\, \SI{0.2}{\cubic\meter\per\second\squared}\right]\\
	&\mathds{P} \left[ h \leq h_\text{max} \right] \geq 0.95 \,.
\end{align}
The stochastic model predictive controller is implemented with GRAMPC-S and moment-based representation with UT, a Gaussian process with 10 data points, and a prediction horizon $T = \SI{1.5}{\second}$. Figure~\ref{fig_gp_res} shows the resulting state and input trajectories for the initial state $q(0) = \SI{0}{\cubic\meter\per\second}$ and $h(0) = \SI{0.1}{\meter}$. In the first case, the data points are generated with a measurement noise variance of $\sigma_\nu^2 = \SI[print-unity-mantissa=false]{e{-3}}{\metre\squared\per\second\squared}$. In the second scenario, the data points are generated with a smaller measurement noise variance of $\sigma_\nu^2 = \SI[print-unity-mantissa=false]{e{-9}}{\metre\squared\per\second\squared}$. The greater measurement noise in the first case results in a higher uncertainty about the behavior of the system and thus to a greater uncertainty of the predicted states. As this uncertainty is explicitly taken into account by the controller, it operates more cautiously by maintaining a larger distance to the constraint.

\begin{figure}
	\centering
	\begin{tikzpicture}
\tikzset{font=\small}

\begin{axis}[%
width=3.97in,
height=1.3in,
scale only axis,
xmode=log,
xlabel={Number of data points},
ylabel={Computation time in \si{\second}},
xmin=1,
xmax=10000,
xminorticks=true,
ymode=log,
ymin=0.0005,
ymax=20,
yminorticks=true,
max space between ticks=20,
]
\addplot [color=blue1, thick, mark=x, mark options={solid, blue1,thick,mark size=2.5pt}, forget plot]
  table[row sep=crcr]{%
1	0.000573\\
10	0.00245\\
100	0.01579\\
1000		0.199\\
10000	10.626\\
};
\end{axis}

\begin{axis}[%
width=4in,
height=1.3in,
scale only axis,
xmin=0,
xmax=1,
ymin=0,
ymax=1,
axis line style={draw=none},
ticks=none,
axis x line*=bottom,
axis y line*=left
]
\end{axis}
\end{tikzpicture}%
	\caption{Average computation time per sampling step of the stochastic model predictive controller for the water tank problem based on moment-based representation with UT using Gaussian process regression.}
	\label{fig_gp_comptime}
\end{figure}
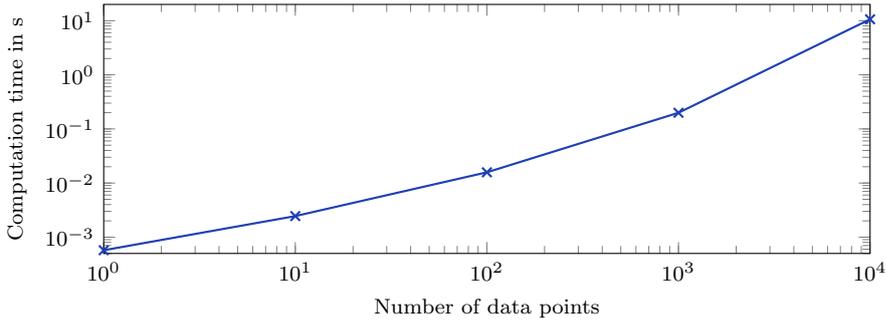

The consideration of a Gaussian process has a large impact on the required computation time, because the evaluation of its mean and covariance matrix involves matrix-vector multiplications, whose dimensions increase with the number of data points. For this reason, the required computation time of the controller per sampling step is evaluated as a function of the number of data points. Figure~\ref{fig_gp_comptime} shows the results for a stochastic model predictive controller for the water tank problem based on MR with UT. The hardware used for the simulations is the same as for the evaluation of $t_\text{CPU}$ in Table~\ref{tab:reactor_compTime}. As Figure~\ref{fig_gp_comptime} shows, the computation time rises significantly as the number of data points increases.

\subsection{Experimental validation} \label{sec:exp_val_inv_pendulum}
\begin{figure}[b]
	\centering
	\includegraphics[scale=0.153]{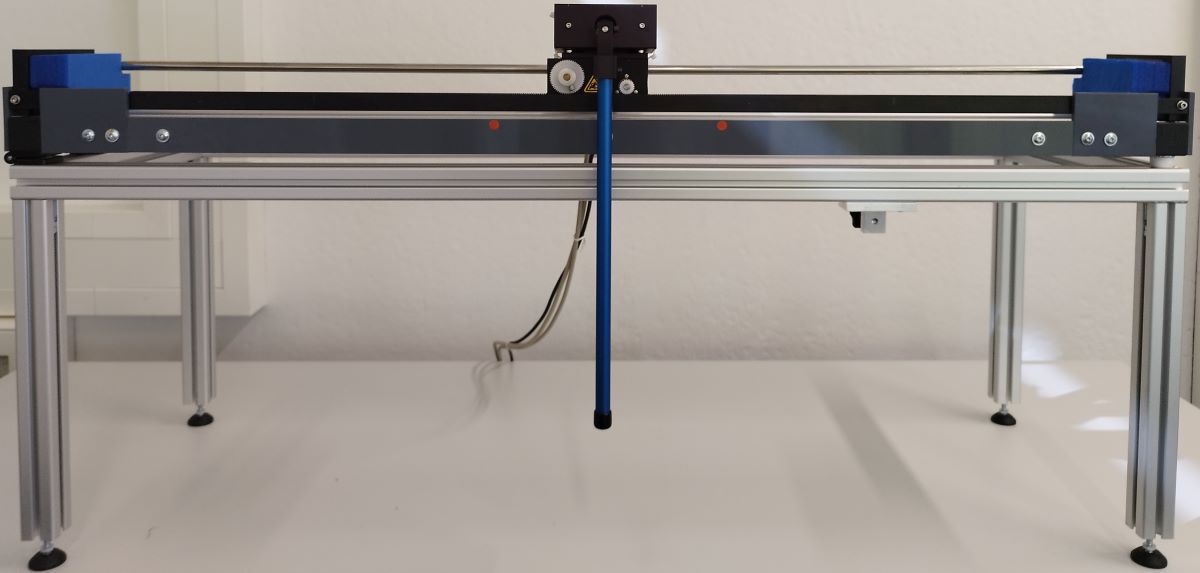}
	\caption{Experimental setup of the inverted pendulum.}
	\label{fig_pendulum}
\end{figure}

To demonstrate its applicability, GRAMPC-S is validated experimentally. For this purpose, an inverse pendulum from Quanser is considered, which is shown in Figure~\ref{fig_pendulum}. It consists of a cart that slides on a track and is controlled by a motor and a pendulum mounted on the cart. The speed of the cart is controlled by a proportional controller. The control input $u$ of the stochastic model predictive controller is the desired velocity of the cart, which is transmitted to the proportional controller. The differential equations for the cart position $x_c$, the cart speed $v_c$, the pendulum angle $\alpha$ and the angular velocity $\omega$ thus result in%
\begin{subequations}%
\begin{align}%
	\dot x_c  &= v_c\\
	\dot v_c &= u\\
	\dot \alpha &= \omega\\
	\dot \omega &= - \frac{d \omega + \frac12 l_p \left( m_p u \cos(\alpha) + g m_p \sin(\alpha) \right)}{\frac14 m_p \l_p^2 + J_p}\,,
\end{align}%
\end{subequations}%
where $d = \SI{2.4e-3}{\newton\metre\second}$ is the damping coefficient, $l_p = \SI{0.356}{\metre}$ is the length of the pendulum, $m_p = \SI{0.127}{\kilogram}$ is the mass of the pendulum, $J_p = \SI{1.198e-3}{\kilogram\metre\squared}$ is the moment of inertia of the pendulum, and $g = \SI{9.81}{\metre\per\second\squared}$ is the gravity acceleration\footnote{Quanser Inc (2012) Linear Flexible Joint with Inverted Pendulum Experiment – User Manual}. The objective of the control is to stabilize the upper equilibrium of the pendulum at $\alpha = 0$. To this end, the cost function
\begin{align}
	J\left( u; \, \vm x_0 \right) = \mathds{E} \left[\int\limits_0^T  100 \left(x_c - x_{c,des}\right)^2 + v_c^2 + \alpha^2 + \omega^2 +  10^{-9}u^2 \,\, \text{d}t \right]
\end{align}
is defined, where $x_{c,des}$ denotes the desired cart position. The uncertainty about the system states arises from the fact that $x_c$ and $\alpha$ can only be measured with limited accuracy. For this reason, a Kalman filter based on the linearized system at the upper equilibrium is used to estimate the mean and covariance matrix of the states. Control and state estimator are implemented on a dSPACE MicroLabBox with a $\SI{2}{\giga\hertz}$ NXP QorlQ P5020 CPU.

\begin{figure}
	\centering
	\input{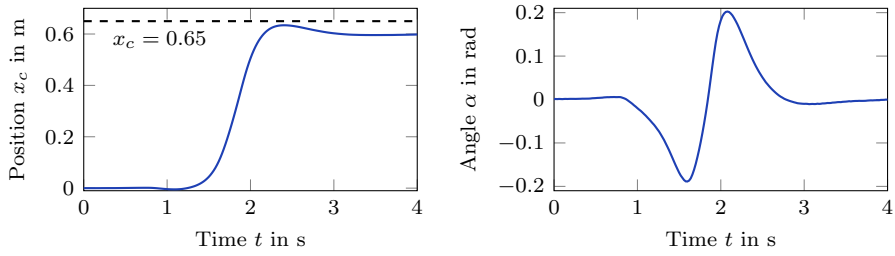}
	\caption{Trajectories of the cart position and the pendulum angle of the controlled system for a setpoint change.}
	\label{fig_eval_pendulum}
\end{figure}

The initial system state is $[x_c,\; v_c, \; \alpha, \; \omega ]\trans = [\SI{0}{\metre},\; \SI{0}{\metre\per\second},\; \SI{0}{\radian},\; \SI{0}{\radian\per\second}]\trans$. The objective of the controller is a setpoint change from $x_{c,des}= \SI{0}{\metre}$ to $x_{c,des}= \SI{0.6}{\metre}$. Furthermore, the controller must satisfy the constraints
\begin{align}
	&\mathds{P} \left[ x_c \leq \SI{0.65}{\metre} \right] \geq 0.95\\
	&u \in \left[ \SI{-5}{\metre\per\second\squared}, \, \SI{5}{\metre\per\second\squared} \right]	\,.
\end{align}
The prediction horizon of the model predictive controller is set to $T = \SI{0.7}{\second}$ and the sampling time is $\Delta t = \SI{1}{\milli\second}$. Figure~\ref{fig_eval_pendulum} shows the resulting trajectories of the cart position and the pendulum angle for SMPC with SR and UT. It can be seen that the controller is able to perform the setpoint change while satisfying the probabilistic constraint. The mean computation time of the stochastic model predictive controller using the dSPACE MicroLabBox is \SI{0.55}{\milli\second}, with a minimum value of \SI{0.53}{\milli\second} and a maximum value of \SI{0.58}{\milli\second}. This example demonstrates that SMPC can be executed in real time for a mechanical system with a sampling time of~\SI{1}{\milli\second} on a rapid control prototyping hardware.

\section{Conclusion} \label{sec:conclusion}
This paper presents the open-source stochastic model predictive control framework GRAMPC-S, which enables the solution of stochastic optimal control problems with probabilistic inequality constraints for nonlinear systems. This may involve uncertainties of the states, the system parameters or in the form of unknown parts of the system dynamics that can be learned by a Gaussian process. One challenge of SMPC is the uncertainty propagation required to compute the stochastic moments of the predicted states. GRAMPC-S provides implementations for uncertainty propagation using linearization with a first-order Taylor series, Stirling interpolation, unscented transformation, Gaussian quadrature, Monte-Carlo method and polynomial chaos expansion. Several options are available to evaluate or conservatively estimate probabilistic constraints based on assumptions on the probability density functions. This allows to reformulate the stochastic optimal control problem as a deterministic one that is solved with the MPC solver GRAMPC. The stochastic properties of the random variables can be represented either by samples or by a finite number of stochastic moments, which leads to different optimization problems and restricts the choice of uncertainty propagation methods. The evaluation shows the applicability of the presented SMPC framework in various technical areas. In particular, the experimental validation demonstrates that the code is computationally efficient and can be applied to control systems with sampling times in the millisecond range. GRAMPC-S is distributed under the Berkeley Software Distribution 3-clause version (BSD-3) license and is available at https://github.com/grampc/grampc-s.

\bibliographystyle{spbasic}
\bibliography{literature}

\end{document}